# Multidimensional Manifold Extraction for Multicarrier Continuous-Variable Quantum Key Distribution

Laszlo Gyongyosi

[1] Quantum Technologies Laboratory, Department of Telecommunications
*Budapest University of Technology and Economics*
2 Magyar tudosok krt, Budapest, *H*-1117, Hungary
[2] MTA-BME Information Systems Research Group
*Hungarian Academy of Sciences*
7 Nador st., Budapest, *H*-1051 Hungary

gyongyosi@hit.bme.hu

**Abstract**

We introduce the multidimensional manifold extraction for multicarrier continuous-variable (CV) quantum key distribution (QKD). The manifold extraction utilizes the resources that are injected into the transmission by the additional degrees of freedom of the multicarrier modulation. We demonstrate the results through the AMQD (adaptive multicarrier quadrature division) scheme, which granulates the information into Gaussian sub-carrier CVs and divides the physical link into several Gaussian sub-channels for the transmission. We prove that the exploitable extra degree of freedom in a multicarrier CVQKD scenario significantly extends the possibilities of single-carrier CVQKD. The manifold extraction allows for the parties to reach decreased error probabilities by utilizing those extra resources of a multicarrier transmission that are not available in a single-carrier CVQKD setting. We define the multidimensional manifold space of multicarrier CVQKD and the optimal tradeoff between the available degrees of freedom of the multicarrier transmission. We also extend the manifold extraction for the multiple-access AMQD-MQA (multiuser quadrature allocation) multicarrier protocol. The additional resources of multicarrier CVQKD allow the achievement of significant performance improvements that are particularly crucial in an experimental scenario.

# 1 Introduction

By utilizing the fundamental laws of quantum mechanics, the continuous-variable quantum key distribution (CVQKD) systems allow to realize an unconditionally secure communication through the currently established communication networks. The CVQKD protocols do not require single photon devices in contrast to the first developed discrete-variable (DV) QKD protocols [1–17]. This significant benefit has immediately made possible to achieve the practical implementation of QKD by the standard devices of traditional telecommunications [18–19], [23–25]. In a CVQKD setting, the information is carried by Gaussian random distributed position and momentum quadratures, which identify a quantum state in the phase space. The quantum states are sent through a noisy link (e.g., an optical fiber or a wireless optical channel [18–19], [26-30]), which adds a white Gaussian noise to the phase space transmission. Despite the fact that the noise characteristic of a CVQKD transmission is plausible and well exploitable in the security proofs, the performance of CVQKD, particularly the currently available secret key rates, is still below the rates of the protocols of traditional telecommunications. This issue brings up a potential requirement on the delivery of an intensive performance enhancement for CVQKD. In particular, for this purpose the *multicarrier* CVQKD modulation has been recently proposed through the multicarrier transmission scheme of AMQD (adaptive multicarrier quadrature division) [2]. The AMQD allows improved secret key rates and higher tolerable excess noise in comparison with standard (referred to as *single-carrier* throughout) CVQKD. The multicarrier transmission granulates the information into several Gaussian subcarrier CVs, which are then transmitted through the Gaussian sub-channels. Particularly, the AMQD divides the physical Gaussian channel into several Gaussian sub-channels; each sub-channel is dedicated for the conveying of a given Gaussian subcarrier CV. Similar to single-carrier CVQKD, a multicarrier CVQKD also provides an unconditional security against the most powerful attacks [2]. Specifically, the benefits of AMQD have been extended by SVD-assistance (singular value decomposition) through the SVD-assisted AMQD [4]. Precisely, the SVD-assistance further injects an additional degree of freedom into the multicarrier transmission. The multicarrier CVQKD has been also proposed for a multiple-access scenario through the AMQD-MQA (multiuser quadrature allocation) scheme [3]. The AMQD-MQA allows for several legal parties to perform reliable simultaneous secret communication over a shared physical Gaussian link through the combination of a sophisticated allocation mechanism of the Gaussian subcarriers and the careful utilization of the Gaussian sub-channels. The secret key rates and security thresholds of multicarrier transmission have been proven in [5], leading to enhanced secret key rates in both one- and two-way CVQKD. The common root of these improvements is that the additional degrees of freedom injected by the multicarrier transmission act as a resource, allowing for the parties to exceed significantly the possibilities of singe-carrier CVQKD. We also further confirm this statement in this work through the utilization of these extra resources.

In traditional communications, the diversity is an effective technique to improve the performance of communication over a noisy channel. The diversity can be obtained through several different tools—most of these solutions are based on sophisticated information coding approaches. The diversity of classical communication channels represents an extra resource from which several benefits can be extracted to improve the performance of the transmission. The diversity in traditional telecommunications can be obtained via time, frequency, space, and coding [20–22]. Basically, in a communication scenario, the diversity provides a useful tool to improve the rates and the reliability. Here we show that similar benefits can be obtained for a multicarrier CVQKD scenario. The proposed solution is called *manifold extraction*. We propose the manifold extraction for multicarrier CVQKD, achieving an improved transmission by utilizing those available additional degrees of freedom in the Gaussian quantum channel that are obviously not available in a single-carrier CVQKD setting. In particular, the extractable manifold provided by the additional degrees of freedom of a multicarrier CVQKD transmission also allows for the parties in a multiple-access scenario to decrease much more significantly the error probabilities than it does presently in a single-carrier scheme. Specifically, the origin of these benefits is that the additional degrees of freedom of the multicarrier CVQKD provide an exploitable resource for the legal parties.

The proposed manifold extraction uses a sophisticated phase space constellation for the Gaussian sub-channels [4] which provides a natural framework to exploit the manifold patterns of the sub-channel transmittance coefficients. The manifold extraction can be applied for an arbitrary distribution of the sub-channel transmittance coefficients and, by exploiting some properties of the phase space constellation it does not require the use of a statistical model. The proposed phase space constellation offers an analogous criterion to an averaging over the statistics of the sub-channel transmittance coefficients. We compare the achievable performance of manifold extraction of multicarrier and single-carrier CVQKD. We determine the optimal *manifold-degree of freedom ratio tradeoff curve* and define its attributes in a single and multicarrier CVQKD setting. We prove that the manifold extraction in a multicarrier scenario offers significantly decreased error probabilities, and through the sophisticated allocation of the Gaussian subcarrier CVs this benefit can be extended to all legal users of a multiple-access multicarrier CVQKD. We characterize the *multidimensional manifold space* of multicarrier CVQKD and define the multidimensional optimal tradeoff function in a high-dimensional manifold space. We then study the manifold extraction for multicarrier CVQKD through AMQD, and multiple-access multicarrier CVQKD through AMQD-MQA, respectively.

This paper is organized as follows. Section 2 summarizes some preliminary findings. Section 3 defines the multidimensional manifold space for CVQKD. Section 4 proposes the manifold extraction of multicarrier CVQKD and multiple-access multicarrier CVQKD. Finally, Section 5 concludes the results. Supplemental Information is included in the Appendix.

## 2 Preliminaries

In Section 2, we briefly summarize the notations and basic terms. For further information, see the detailed descriptions of [2–5].

## 2.1 Basic Terms and Definitions

### 2.1.1 Multicarrier CVQKD

In this section we very briefly summarize the basic notations of AMQD from [2]. The following description assumes a single user, and the use of $n$ Gaussian sub-channels $\mathcal{N}_i$ for the transmission of the subcarriers, from which only $l$ sub-channels will carry valuable information.

In the single-carrier modulation scheme, the $j$-th input single-carrier state $|\varphi_j\rangle = |x_j + \mathrm{i} p_j\rangle$ is a Gaussian state in the phase space $\mathcal{S}$, with i.i.d. Gaussian random position and momentum quadratures $x_j \in \mathbb{N}\left(0, \sigma_{\omega_0}^2\right)$, $p_j \in \mathbb{N}\left(0, \sigma_{\omega_0}^2\right)$, where $\sigma_{\omega_0}^2$ is the modulation variance of the quadratures. In the multicarrier scenario, the information is carried by Gaussian subcarrier CVs, $|\phi_i\rangle = |x_i + \mathrm{i} p_i\rangle$, $x_i \in \mathbb{N}\left(0, \sigma_\omega^2\right)$, $p_i \in \mathbb{N}\left(0, \sigma_\omega^2\right)$, where $\sigma_\omega^2$ is the modulation variance of the subcarrier quadratures, which are transmitted through a noisy Gaussian sub-channel $\mathcal{N}_i$. Precisely, each $\mathcal{N}_i$ Gaussian sub-channel is dedicated for the transmission of one Gaussian subcarrier CV from the $n$ subcarrier CVs. (*Note*: index $l$ refers to the subcarriers, while index $j$, to the single-carriers, throughout the manuscript.) The single-carrier state $|\varphi_j\rangle$ in the phase space $\mathcal{S}$ can be modeled as a zero-mean, circular symmetric complex Gaussian random variable $z_j \in \mathcal{CN}\left(0, \sigma_{\omega_{z_j}}^2\right)$, with variance $\sigma_{\omega_{z_j}}^2 = \mathbb{E}\left[\left|z_j\right|^2\right]$, and with i.i.d. real and imaginary zero-mean Gaussian random components $\mathrm{Re}\left(z_j\right) \in \mathbb{N}\left(0, \sigma_{\omega_0}^2\right)$, $\mathrm{Im}\left(z_j\right) \in \mathbb{N}\left(0, \sigma_{\omega_0}^2\right)$.

In the multicarrier CVQKD scenario, let $n$ be the number of Alice's input single-carrier Gaussian states. Precisely, the $n$ input coherent states are modeled by an $n$-dimensional, zero-mean, circular symmetric complex random Gaussian vector

$$\mathbf{z} = \mathbf{x} + \mathrm{i}\mathbf{p} = \left(z_1, \ldots, z_n\right)^T \in \mathcal{CN}\left(0, \mathbf{K_z}\right), \tag{1}$$

where each $z_j$ is a zero-mean, circular symmetric complex Gaussian random variable

$$z_j \in \mathcal{CN}\left(0, \sigma_{\omega_{z_j}}^2\right), \ z_j = x_j + \mathrm{i} p_j. \tag{2}$$

Specifically, the real and imaginary variables (i.e., the position and momentum quadratures) formulate $n$-dimensional real Gaussian random vectors, $\mathbf{x} = \left(x_1, \ldots, x_n\right)^T$ and $\mathbf{p} = \left(p_1, \ldots, p_n\right)^T$, with zero-mean Gaussian random variables

$$f\left(x_j\right) = \frac{1}{\sigma_{\omega_0}\sqrt{2\pi}} e^{\frac{-x_j^2}{2\sigma_{\omega_0}^2}}, \ f\left(p_j\right) = \frac{1}{\sigma_{\omega_0}\sqrt{2\pi}} e^{\frac{-p_j^2}{2\sigma_{\omega_0}^2}}, \tag{3}$$

where $\mathbf{K_z}$ is the $n \times n$ Hermitian covariance matrix of $\mathbf{z}$:

$$\mathbf{K_z} = \mathbb{E}\left[\mathbf{z}\mathbf{z}^\dagger\right], \tag{4}$$

while $\mathbf{z}^\dagger$ is the adjoint of $\mathbf{z}$. For vector $\mathbf{z}$, $\mathbb{E}[\mathbf{z}] = \mathbb{E}[e^{i\gamma}\mathbf{z}] = \mathbb{E}e^{i\gamma}[\mathbf{z}]$ holds, and

$$\mathbb{E}[\mathbf{z}\mathbf{z}^T] = \mathbb{E}\left[e^{i\gamma}\mathbf{z}\left(e^{i\gamma}\mathbf{z}\right)^T\right] = \mathbb{E}e^{i2\gamma}[\mathbf{z}\mathbf{z}^T], \tag{5}$$

for any $\gamma \in [0, 2\pi]$. The density of $\mathbf{z}$ is as follows (assuming that $\mathbf{K}_\mathbf{z}$ is invertible):

$$f(\mathbf{z}) = \frac{1}{\pi^n \det \mathbf{K}_\mathbf{z}} e^{-\mathbf{z}^\dagger \mathbf{K}_\mathbf{z}^{-1} \mathbf{z}}. \tag{6}$$

A $n$-dimensional Gaussian random vector is expressed as $\mathbf{x} = \mathbf{As}$, where $\mathbf{A}$ is an (invertible) linear transform from $\mathbb{R}^n$ to $\mathbb{R}^n$, and $\mathbf{s}$ is an $n$-dimensional standard Gaussian random vector $\mathbb{N}(0,1)_n$. This vector is characterized by its covariance matrix $\mathbf{K}_\mathbf{x} = \mathbb{E}[\mathbf{x}\mathbf{x}^T] = \mathbf{A}\mathbf{A}^T$, as

$$\mathbf{x} = \frac{1}{\left(\sqrt{2\pi}\right)^n \sqrt{\det(\mathbf{A}\mathbf{A}^T)}} e^{-\frac{\mathbf{x}^T \mathbf{x}}{2(\mathbf{A}\mathbf{A}^T)}}. \tag{7}$$

The Fourier transformation $F(\cdot)$ of the $n$-dimensional Gaussian random vector $\mathbf{v} = (v_1, \ldots, v_n)^T$ results in the $n$-dimensional Gaussian random vector $\mathbf{m} = (m_1, \ldots, m_n)^T$, precisely:

$$\mathbf{m} = F(\mathbf{v}) = e^{\frac{-\mathbf{m}^T \mathbf{A}\mathbf{A}^T \mathbf{m}}{2}} = e^{\frac{-\sigma_{\omega_0}^2 (m_1^2 + \ldots + m_n^2)}{2}}. \tag{8}$$

In the first step of AMQD, Alice applies the inverse FFT (fast Fourier transform) operation to vector $\mathbf{z}$ (see (1)), which results in an $n$-dimensional zero-mean, circular symmetric complex Gaussian random vector $\mathbf{d}$, $\mathbf{d} \in \mathcal{CN}(0, \mathbf{K}_\mathbf{d})$, $\mathbf{d} = (d_1, \ldots, d_n)^T$, precisely as

$$\mathbf{d} = F^{-1}(\mathbf{z}) = e^{\frac{\mathbf{d}^T \mathbf{A}\mathbf{A}^T \mathbf{d}}{2}} = e^{\frac{\sigma_{\omega_0}^2 (d_1^2 + \ldots + d_n^2)}{2}}, \tag{9}$$

where

$$d_i = x_{d_i} + i p_{d_i}, \; d_i \in \mathcal{CN}(0, \sigma_{d_i}^2), \tag{10}$$

where $\sigma_{\omega_{d_i}}^2 = \mathbb{E}[|d_i|^2]$ and the position and momentum quadratures of $|\phi_i\rangle$ are i.i.d. Gaussian random variables

$$\operatorname{Re}(d_i) = x_{d_i} \in \mathbb{N}(0, \sigma_{\omega_i}^2), \; \operatorname{Im}(d_i) = p_{d_i} \in \mathbb{N}(0, \sigma_{\omega_i}^2), \tag{11}$$

where $\mathbf{K}_\mathbf{d} = \mathbb{E}[\mathbf{d}\mathbf{d}^\dagger]$, $\mathbb{E}[\mathbf{d}] = \mathbb{E}[e^{i\gamma}\mathbf{d}] = \mathbb{E}e^{i\gamma}[\mathbf{d}]$, and $\mathbb{E}[\mathbf{d}\mathbf{d}^T] = \mathbb{E}\left[e^{i\gamma}\mathbf{d}\left(e^{i\gamma}\mathbf{d}\right)^T\right] = \mathbb{E}e^{i2\gamma}[\mathbf{d}\mathbf{d}^T]$ for any $\gamma \in [0, 2\pi]$.

The $\mathbf{T}(\mathcal{N})$ transmittance vector of $\mathcal{N}$ in the multicarrier transmission is

$$\mathbf{T}(\mathcal{N}) = [T_1(\mathcal{N}_1), \ldots, T_n(\mathcal{N}_n)]^T \in \mathcal{C}^n, \tag{12}$$

where
$$T_i(\mathcal{N}_i) = \mathrm{Re}(T_i(\mathcal{N}_i)) + \mathrm{i}\,\mathrm{Im}(T_i(\mathcal{N}_i)) \in \mathcal{C}, \tag{13}$$

is a complex variable, which quantifies the position and momentum quadrature transmission (i.e., gain) of the $i$-th Gaussian sub-channel $\mathcal{N}_i$, in the phase space $\mathcal{S}$, with real and imaginary parts
$$0 \leq \mathrm{Re}\,T_i(\mathcal{N}_i) \leq 1/\sqrt{2}, \text{ and } 0 \leq \mathrm{Im}\,T_i(\mathcal{N}_i) \leq 1/\sqrt{2}. \tag{14}$$

Particularly, the $T_i(\mathcal{N}_i)$ variable has the squared magnitude of
$$\left|T_i(\mathcal{N}_i)\right|^2 = \mathrm{Re}\,T_i(\mathcal{N}_i)^2 + \mathrm{Im}\,T_i(\mathcal{N}_i)^2 \in \mathbb{R}, \tag{15}$$

where
$$\mathrm{Re}\,T_i(\mathcal{N}_i) = \mathrm{Im}\,T_i(\mathcal{N}_i). \tag{16}$$

The Fourier-transformed transmittance of the $i$-th sub-channel $\mathcal{N}_i$ (resulted from CVQFT (continuous-variable quantum Fourier transform) operation at Bob) is denoted by
$$\left|F(T_i(\mathcal{N}_i))\right|^2. \tag{17}$$

The $n$-dimensional zero-mean, circular symmetric complex Gaussian noise vector $\Delta \in \mathcal{CN}(0, \sigma_\Delta^2)_n$ of the quantum channel $\mathcal{N}$, is evaluated as
$$\Delta = (\Delta_1, \ldots, \Delta_n)^T \in \mathcal{CN}(0, \mathbf{K}_\Delta), \tag{18}$$

where
$$\mathbf{K}_\Delta = \mathbb{E}[\Delta \Delta^\dagger], \tag{19}$$

with independent, zero-mean Gaussian random components
$$\Delta_{x_i} \in \mathbb{N}(0, \sigma_{\mathcal{N}_i}^2), \text{ and } \Delta_{p_i} \in \mathbb{N}(0, \sigma_{\mathcal{N}_i}^2), \tag{20}$$

with variance $\sigma_{\mathcal{N}_i}^2$, for each $\Delta_i$ of a Gaussian sub-channel $\mathcal{N}_i$, which identifies the Gaussian noise of the $i$-th sub-channel $\mathcal{N}_i$ on the quadrature components in the phase space $\mathcal{S}$.

The CVQFT-transformed noise vector can be rewritten as
$$F(\Delta) = (F(\Delta_1), \ldots, F(\Delta_n))^T, \tag{21}$$

with independent components $F(\Delta_{x_i}) \in \mathbb{N}(0, \sigma_{F(\mathcal{N}_i)}^2)$ and $F(\Delta_{p_i}) \in \mathbb{N}(0, \sigma_{F(\mathcal{N}_i)}^2)$ on the quadratures, for each $F(\Delta_i)$. Precisely, it also defines an $n$-dimensional zero-mean, circular symmetric complex Gaussian random vector $F(\Delta) \in \mathcal{CN}(0, \mathbf{K}_{F(\Delta)})$ with a covariance matrix
$$\mathbf{K}_{F(\Delta)} = \mathbb{E}\left[F(\Delta)F(\Delta)^\dagger\right], \tag{22}$$

where $\mathbf{K}_{F(\Delta)} = \mathbf{K}_\Delta$, by theory.

### 2.1.2 SVD-Assisted Multicarrier CVQKD

We briefly summarize the SVD-assisted multicarrier CVQKD scheme from [4]. The detailed description of the multiple-access AMQD-MQA scheme is included in [3].

Precisely, the singular layer consists of a pre-unitary $F_1$ ($U_1$) (scaled FFT operation (scaled CVQFT), independent from the IFFT (inverse Fast Fourier transform) operation $F$ ($U_1$)) and a post-unitary $U_2^{-1}$ (CVQFT operation, independent from the $U$ CVQFT$^\dagger$ operation) that perform the pre- and post-transform.

The pre-unitary $F_1$ ($U_1$) transforms such that the input will be sent through the $\lambda_i$ eigenchannels of the Gaussian link, whereas $U_2^{-1}$ performs its inverse. Note that the pre-$F_1$ ($U_1$) and post-$U_2^{-1}$ unitaries are the not inverse of $F$ and $U$ but $F_1^{-1}$ ($U_1^{-1}$) and $U_2$, respectively. In particular, these unitaries define the set $S_1$ of singular operators, as follows [4]:

$$S_1 = \{F_1, U_2^{-1}\}. \tag{23}$$

Specifically, if each transmit user sends a single-carrier Gaussian CV signal to an encoder $\mathcal{E}$, then the pre-operator is the unitary $U_1$, the CVQFT operation, whereas the unitary post-operator is achieved by the inverse CVQFT operation $U_2^{-1}$, defining the set $S_2$ of singular operators as

$$S_2 = \{U_1, U_2^{-1}\}. \tag{24}$$

The subindices of the operators $\{F_1, U_2^{-1}\}$ and $\{U_1, U_2^{-1}\}$ are different in each $S_i, i = 1, 2$ because these operators are not the inverse of each other.

These operators are determined by the SVD of $F(\mathbf{T})$, which is evaluated as

$$F(\mathbf{T}) = U_2 \Gamma F_1^{-1}, \tag{25}$$

where $F_1^{-1}, F_1 \in \mathbb{C}^{K_{in} \times K_{in}}$ and $U_2, U_2^{-1} \in \mathbb{C}^{K_{out} \times K_{out}}$, $K_{in}$, and $K_{out}$ refer to the number of sender and receiver users such that

$$K_{in} \leq K_{out}, \ F_1^{-1} F_1 = F_1 F_1^{-1} = I, \tag{26}$$

and

$$U_2 U_2^{-1} = U_2^{-1} U_2 = I. \tag{27}$$

The term $\Gamma \in \mathbb{R}$ is a diagonal matrix with nonnegative real diagonal elements

$$\lambda_1 \geq \lambda_2 \geq \ldots \lambda_{n_{\min}}, \tag{28}$$

which are called the *eigenchannels* of $F(\mathbf{T}) = U_2 \Gamma F_1^{-1}$, where

$$n_{\min} = \min(K_{in}, K_{out}), \tag{29}$$

which equals to the rank of $F(\mathbf{T})$, where

$$K_{in} \leq K_{out}, \tag{30}$$

by an initial assumption. (*Note*: the eigenchannels are also called the ordered singular values of $F(\mathbf{T})$.) In terms of the $\lambda_i$ eigenchannels, $F(\mathbf{T})$ can be precisely rewritten as

$$F(\mathbf{T}) = \sum\nolimits_{n_{\min}} \lambda_i U_{2,i} F_{1,i}^{-1}, \qquad (31)$$

where $\lambda_i U_{2,i} F_{1,i}^{-1}$ are rank-one matrices. In fact, the $n_{\min}$ squared eigenchannels $\lambda_i^2$ are the eigenvalues of the matrix

$$F(\mathbf{T}) F(\mathbf{T})^{\dagger} = U_2 \Gamma \Gamma^T U_2^{-1}, \qquad (32)$$

where $\Gamma^T$ is the transpose of $\Gamma$.

The complete description of the singular layer of CVQKD can be found in [4].

### 2.1.3 Rate Formulas of Multicarrier CVQKD

The complete derivation of the secret key rate formulas can be found in [5], here we give a brief overview on the transmission rates of multicarrier CVQKD.

In particular, the (real domain) classical capacity of a Gaussian sub-channel $\mathcal{N}_i$ in the multicarrier setting is

$$C(\mathcal{N}_i) = \tfrac{1}{2} \log_2 \left( 1 + \tfrac{\sigma_{\omega_i}^2 |F(T_i(\mathcal{N}_i))|^2}{\sigma_{\mathcal{N}_i}^2} \right), \qquad (33)$$

while in the SVD-assisted AMQD,

$$C'(\mathcal{N}_i) = \tfrac{1}{2} \log_2 \left( 1 + \tfrac{\sigma_{\omega_i''}^2 |F(T_i(\mathcal{N}_i))|^2}{\sigma_{\mathcal{N}_i}^2} \right), \qquad (34)$$

where $\sigma_{\omega''}^2 = \sigma_{\omega}^2 (1 + c) > \sigma_{\omega}^2$.

Specifically, the SNR (signal to noise ratio) of $\mathcal{N}_i$ is expressed as

$$\mathrm{SNR}_i = \tfrac{\sigma_{\omega_i}^2}{\sigma_{\mathcal{N}_i}^2}, \qquad (35)$$

while the SNR of $\mathcal{N}$ at a constant modulation variance $\sigma_{\omega}^2$ is $\mathrm{SNR} = \tfrac{\sigma_{\omega}^2}{\sigma_{\mathcal{N}}^2}$.

Particularly, in the SVD-assisted AMQD, it referred to as

$$\mathrm{SNR}_i' = \tfrac{\sigma_{\omega_i''}^2}{\sigma_{\mathcal{N}_i}^2}, \text{ and } \mathrm{SNR}' = \tfrac{\sigma_{\omega''}^2}{\sigma_{\mathcal{N}}^2}, \qquad (36)$$

respectively. From (33) and (34), the (real domain) classical information transmission rates $R_k(\mathcal{N})$ and $R_k'(\mathcal{N})$ of user $U_k$ through the $l$ $\mathcal{N}_i$ Gaussian sub-channels in AMQD and SVD-assisted AMQD are precisely as follows:

$$R_k(\mathcal{N}) \leq \max_{\forall i} \mathbb{E} \left[ \sum\nolimits_l \tfrac{1}{2} \log_2 \left( 1 + \tfrac{\sigma_{\omega_i}^2 |F(T_i(\mathcal{N}_i))|^2}{\sigma_{\mathcal{N}}^2} \right) \right], \qquad (37)$$

and

$$R_k'(\mathcal{N}) \leq \max_{\forall i} \mathbb{E} \left[ \sum\nolimits_l \tfrac{1}{2} \log_2 \left( 1 + \tfrac{\sigma_{\omega_i''}^2 |F(T_i(\mathcal{N}_i))|^2}{\sigma_{\mathcal{N}}^2} \right) \right]. \qquad (38)$$

Precisely, the $\text{SNR}_i^*$ (signal to noise ratio) of the $i$-th Gaussian sub-channel $\mathcal{N}_i$ for the transmission of private classical information (i.e., for the derivation of the secret key rate) under an optimal Gaussian attack [5], [14-16] is expressed as

$$\text{SNR}_i^* = \frac{\sigma_{\omega_i}^2}{\sigma_{\mathcal{N}_i^*}^2}, \text{ and } \text{SNR}^* = \frac{\sigma_{\omega}^2}{\sigma_{\mathcal{N}^*}^2}, \tag{39}$$

where $\sigma_{\mathcal{N}_i^*}^2$ is precisely evaluated as [5]

$$\sigma_{\mathcal{N}_i^*}^2 = \sigma_{\omega_i}^2 \left( \frac{\sigma_{\omega_i}^2 |F(T_i(\mathcal{N}_i))|^2 + \sigma_{X_i}^2}{1 + \sigma_{X_i}^2 \sigma_{\omega_i}^2 |F(T_i(\mathcal{N}_i))|^2} - 1 \right)^{-1}, \tag{40}$$

where

$$\sigma_{X_i}^2 = \sigma_0^2 + N_i, \tag{41}$$

and where $\sigma_0^2$ is the vacuum noise and $N_i$ is the excess noise of the Gaussian sub-channel $\mathcal{N}_i$ defined as

$$N_i = \frac{(W_i - 1)\left( |F(T_{Eve,i})|^2 \right)}{1 - |F(T_{Eve,i})|^2}, \tag{42}$$

where $W_i$ is the variance of Eve's EPR state used for the attacking of $\mathcal{N}_i$, while

$$|T_{Eve,i}|^2 = 1 - |T_i|^2 \tag{43}$$

is the transmittance of Eve's beam splitter (BS), and $|T_i|^2$ is the transmittance of $\mathcal{N}_i$.

Precisely, in the SVD-assisted multicarrier CVQKD,

$$\left(\text{SNR}_i'\right)^* = \frac{\sigma_{\omega_i''}^2}{\sigma_{\mathcal{N}_i^*}^2} \text{ and } \left(\text{SNR}'\right)^* = \frac{\sigma_{\omega''}^2}{\sigma_{\mathcal{N}^*}^2}, \tag{44}$$

for $\mathcal{N}_i$ and $\mathcal{N}$, respectively.

Particularly, from (40) the $P(\mathcal{N}_i)$ private classical capacity (real domain) is expressed as

$$P(\mathcal{N}_i) = \tfrac{1}{2} \log_2 \left( 1 + \frac{\sigma_{\omega_i}^2 |F(T_i(\mathcal{N}_i))|^2}{\sigma_{\mathcal{N}_i^*}^2} \right). \tag{45}$$

The SVD-assisted $P'(\mathcal{N}_i)$ from (44) is then yielded precisely as

$$P'(\mathcal{N}_i) = \tfrac{1}{2} \log_2 \left( 1 + \frac{\sigma_{\omega_i''}^2 |F(T_i(\mathcal{N}_i))|^2}{\sigma_{\mathcal{N}_i^*}^2} \right). \tag{46}$$

Assuming $l$ Gaussian sub-channels, the (real domain) secret key rate $S(\mathcal{N})$ of AMQD and $S'(\mathcal{N})$ of SVD-assisted AMQD are as follows:

$$S(\mathcal{N}) \leq P(\mathcal{N}) = \max_{\forall i} \mathbb{E}\left[ \sum_l \log_2 \left( 1 + \frac{\sigma_{\omega_i}^2 |F(T_i(\mathcal{N}_i))|^2}{\sigma_{\mathcal{N}_i^*}^2} \right) \right], \tag{47}$$

$$S'(\mathcal{N}) \leq P'(\mathcal{N}) = \max_{\forall i} \mathbb{E}\left[ \sum_l \log_2 \left( 1 + \frac{\sigma_{\omega_i''}^2 |F(T_i(\mathcal{N}_i))|^2}{\sigma_{\mathcal{N}_i^*}^2} \right) \right]. \tag{48}$$

## 2.2 Manifold Extraction

In a multicarrier CVQKD scenario, the term *manifold* is interpreted as follows. Let the $i$-th component $p_{j,i}$ of a given private random codeword $\mathbf{p}_j = \left( p_{j,1},\ldots,p_{j,l} \right)^T$ to be transmitted through $\mathcal{N}_i$, where each Gaussian sub-channel is characterized by an independent transmittance coefficient $\left| \left( T_i \left( \mathcal{N}_i \right) \right) \right|^2$. As a first approach, the number $l$ of the Gaussian sub-channels is identified as the *manifold* of $\mathcal{N}$. Precisely, the information is granulated into subcarriers, which are dispersed by the inverse Fourier transform, and each $p_{j,i}$ component is identified by independent transmittance coefficients. A more detailed formula will be concluded in the further sections.

Specifically, the transmission can be utilized by a permutation phase space constellation $\mathcal{C}_\mathcal{S}^P \left( \mathcal{N} \right)$ [4], which was recently proposed for SVD-assisted CVQKD in [4]. Using $P_i$, $i=2,\ldots,l$ random permutation operators, $\mathcal{C}_\mathcal{S}^P \left( \mathcal{N} \right)$ can be defined for the multicarrier transmission as

$$\begin{aligned} \mathcal{C}_\mathcal{S}^P \left( \mathcal{N} \right) &= \left( \mathcal{C}_\mathcal{S} \left( \mathcal{N}_1 \right),\ldots,\mathcal{C}_\mathcal{S} \left( \mathcal{N}_l \right) \right) \\ &= \left( \mathcal{C}_\mathcal{S} \left( \mathcal{N}_1 \right), P_2 \mathcal{C}_\mathcal{S} \left( \mathcal{N}_1 \right),\ldots, P_l \mathcal{C}_\mathcal{S} \left( \mathcal{N}_1 \right) \right), \end{aligned} \quad (49)$$

where $d_{\mathcal{C}_\mathcal{S}(\mathcal{N}_i)} = d_{\mathcal{C}_\mathcal{S}(\mathcal{N}_j)}$ is the cardinality of $\mathcal{C}_\mathcal{S}\left( \mathcal{N}_i \right)$. Using $\mathcal{C}_\mathcal{S}^P \left( \mathcal{N} \right)$, the available degrees of freedom in the Gaussian link can be utilized, and the random permutation operators inject correlation between the $\mathcal{N}_i$ sub-channels via $P_i \mathcal{C}_\mathcal{S}\left( \mathcal{N}_1 \right)$.

In particular, for each Gaussian sub-channel, the distance between the phase space constellation points is evaluated by $\delta_i$, the normalized difference function. Assuming two $l$-dimensional input random private codewords $\mathbf{p}_A = \left( p_{A,1},\ldots p_{A,l} \right)^T$ and $\mathbf{p}_B = \left( p_{B,1},\ldots p_{B,l} \right)^T$ and two Gaussian sub-channels $\mathcal{N}_i$ and $\mathcal{N}_j$, $\delta_i$ is calculated precisely as follows:

$$\delta_i = \frac{1}{\sqrt{\frac{\sigma_{\omega_n}^2}{\sigma_{\mathcal{N}^*}^2}}} \left( p_{A,i} - p_{B,i} \right), \quad (50)$$

particularly, for the $l$ Gaussian sub-channels

$$\left| \delta_{1\ldots l} \right|^2 > \left( c \frac{1}{l 2^{S'(\mathcal{N}_i)}} \right)^l, \quad (51)$$

where the term $\left| \delta_{1\ldots l} \right|$ is referred to as the *product distance* [20-22]. The maximization of this term ensures the maximization of the extractable manifold, and determines the $\tilde{p}_{err}$ pairwise worst-case error probabilities of $\mathbf{p}_A, \mathbf{p}_B$.

As we show in Section 3, by using $\mathcal{C}_\mathcal{S}^P \left( \mathcal{N} \right)$ and (51), the $\tilde{p}_{err}$ worst-case pairwise error probability can be decreased to the theoretical lower bound. We further reveal that in a multiuser CVQKD scenario, this condition can be extended simultaneously for all uses.

Let us assume that the $S_k'\left( \mathcal{N} \right)$ secret key rate of $U_k$, for $\forall k$, is fixed precisely as follows:

$$S_k'\left( \mathcal{N} \right) = \frac{c_k}{n_{\min}} P'\left( \mathcal{N} \right), \quad (52)$$

where $\varsigma_k > 0$ is referred to as the *degree of freedom ratio* of $U_k$, and $n_{\min}$ has been shown in (29). As one can immediately conclude from (52), $S'_k(\mathcal{N}) \ll P'(\mathcal{N})$.

Without loss of generality, for a given sub-channel $\mathcal{N}_i$, we redefine $S'_k(\mathcal{N}_i)$, $\varsigma_{k,i} > 0$ precisely as

$$S'_k(\mathcal{N}_i) = \frac{\varsigma_{k,i}}{n_{\min}} P'(\mathcal{N}_i). \tag{53}$$

(*Note*: From this point, we use the complex domain formulas throughout the manuscript and $S'_k(\mathcal{N})$ and $S'_k(\mathcal{N}_i)$ are fixed to (52) and (53).)

For a given $\mathcal{N}_i$, an $\mathrm{E}_{err}$ error event [20–22] is identified as follows:

$$\mathrm{E}_{err} \equiv \log_2\left(1 + \left|F(T_i(\mathcal{N}_i))\right|^2 (\mathrm{SNR}'_i)^*\right) < S'(\mathcal{N}_i), \tag{54}$$

and the probability of $\mathrm{E}_{err}$ at a given $S'(\mathcal{N}_i)$ is identified by the $p_{err}$ error probability as follows:

$$\mathrm{E}_{err} = p_{err}(S'_k(\mathcal{N}_i)) = \Pr\left(\log_2\left(1 + \left|F(T_i(\mathcal{N}_i))\right|^2 (\mathrm{SNR}'_i)^*\right) < S'(\mathcal{N}_i)\right). \tag{55}$$

Particularly, by some fundamental argumentations on the statistical properties of a Gaussian random distribution [20–22], for $\left|F(T_i(\mathcal{N}_i))\right|^2 (\mathrm{SNR}'_i)^* \to 0$, $p_{err}(S'_k(\mathcal{N}_i))$ can be expressed as

$$p_{err}(S'_k(\mathcal{N}_i)) = \Pr\left(\left|F(T_i(\mathcal{N}_i))\right|^2 (\mathrm{SNR}'_i)^* \log_2 e < S'(\mathcal{N}_i)\right), \tag{56}$$

while for $\left|F(T_i(\mathcal{N}_i))\right|^2 (\mathrm{SNR}'_i)^* \to \infty$, the corresponding error probability is as

$$p_{err}(S'_k(\mathcal{N}_i)) = \Pr\left(\log_2\left(\left|F(T_i(\mathcal{N}_i))\right|^2 (\mathrm{SNR}'_i)^*\right) < S'(\mathcal{N}_i)\right). \tag{57}$$

Let $l = 1$, that is, let's consider a single-carrier CVQKD, with $\left|F(T(\mathcal{N}))\right|^2$ of $\mathcal{N}$, with a secret key rate $S'(\mathcal{N})$. In this setting, $p_{err}$ is expressed precisely as [20]

$$\begin{aligned} p_{err}^{single}(S'_k(\mathcal{N})) &= \Pr\left(\log_2\left(1 + \left|F(T(\mathcal{N}))\right|^2 (\mathrm{SNR}')^*\right) < S'(\mathcal{N})\right) \\ &= \Pr\left(\left|F(T(\mathcal{N}))\right|^2 < \frac{1}{(\mathrm{SNR}')^*}\right) \\ &= \frac{1}{(\mathrm{SNR}')^*}, \end{aligned} \tag{58}$$

by theory.

Specifically, assuming a multicarrier CVQKD scenario with $l$ Gaussian sub-channels and secret key rate $S'(\mathcal{N}_i)$ per $\mathcal{N}_i$, $p_{err}^{AMQD}$ is derived as follows:

Without loss of generality, we construct the set $\mathcal{T}$, such that

$$\mathcal{T}: \min_{\forall i}\left\{\left|F(T_i(\mathcal{N}_i))\right|\right\}, \tag{59}$$

where for $\forall i, i = 1,\ldots l$ the following condition holds:

$$F(T_i(\mathcal{N}_i)) \geq \frac{1}{(\mathrm{SNR}')^*}. \tag{60}$$

In particular, the transmission through the Gaussian sub-channels is evaluated via set $\mathcal{T}$, which refers to the worst-case scenario at which a $S'(\mathcal{N}) > 0$ nonzero secret key rate is possible, by convention. Particularly, in (51), a given $\partial_i$ identifies the minimum distance between the normalized $2^{S'_k(\mathcal{N}_i)}$ points for the phase space constellation $\mathcal{C}'_{\mathcal{S}}(\mathcal{N}_i)$ of $\mathcal{N}_i$.

Precisely, by fundamental theory [20-22], it can be proven that for an arbitrary distribution of the $F(T_i(\mathcal{N}_i))$ Fourier transformed transmittance coefficient, the maximized product distance function of (51) can be derived by an averaging over the following statistic $\mathcal{S}$:

$$\mathcal{S}: F(T_i(\mathcal{N}_i)) \in \mathcal{CN}\left(0, \sigma^2_{F(T_i(\mathcal{N}_i))}\right), \tag{61}$$

where $\sigma^2_{F(T_i(\mathcal{N}_i))} = \mathbb{E}\left[\left|F(T_i(\mathcal{N}_i))\right|^2\right]$, and $F(T_i(\mathcal{N}_i))$ is a zero-mean, circular symmetric complex Gaussian random variable with i.i.d. $\mathbb{N}\left(0, 0.5\sigma^2_{F(T_i(\mathcal{N}_i))}\right)$ zero-mean Gaussian random variables per quadrature components $x_i$ and $p_i$, for the $i$-th Gaussian subcarrier CV.

Putting the pieces together, the maximized product distance function $\left|\delta_{1\ldots l}\right|$ of (51) precisely can be obtained via an averaging over the $\mathcal{S}$ statistics of (61); however, (61) is, in fact, strictly provides an analogous criteria of the worst-case $\tilde{p}_{err}$ situation in (59) via a sophisticated phase space constellation $\mathcal{C}_{\mathcal{S}}$, by theory [20–21]. In other words, set $\mathcal{T}$, as it is given in (59) together with $\mathcal{C}_{\mathcal{S}}$ represents a universal criteria and provides us an alternative solution to find the worst-case $\tilde{p}_{err}$ error probability for arbitrary distributed $F(T_i(\mathcal{N}_i))$ coefficients in a multicarrier CVQKD scenario. Specifically, some of these argumentations can be further exploited in our analysis.

First of all, by using (61), the averaged term $\frac{1}{l}\sum_l \left|F(T_i(\mathcal{N}_i))\right|^2$ can be modeled as a sum of $\mathcal{CN}\left(0, \sigma^2_{F(T_i(\mathcal{N}_i))}\right)$ distributed random variables, with zero mean and variance of $\sigma^2_{F(T_i(\mathcal{N}_i))}$ for each $\mathcal{N}_i$ sub-channels. Then, since $\frac{1}{l}\sum_l \left|F(T_i(\mathcal{N}_i))\right|^2$ is the averaged sum of $2l$ independent real Gaussian random variables, the distribution of $\frac{1}{l}\sum_l \left|F(T_i(\mathcal{N}_i))\right|^2$ precisely can be approximated by a $\chi^2_{2l}$ chi-square distribution with $2l$ degrees of freedom, by a density function $f(\cdot)$:

$$f(x) = \frac{1}{(l-1)!} x^{l-1} e^{-x}, \tag{62}$$

where $x \geq 0$.

In particular, for $x \to 0$, the density can be written as

$$f(x) \approx \frac{1}{(l-1)!} x^{l-1}. \tag{63}$$

Thus, we arrive at $p_{err}^{AMQD}$ as

$$p_{err}^{AMQD} = \Pr\left(\tfrac{1}{l}\sum_l |F(T_i(\mathcal{N}_i))|^2 < \tfrac{1}{(\text{SNR}')^*}\right)$$

$$= \int_0^{\frac{1}{\left[(\text{SNR}')^*\right]}} \tfrac{1}{(l-1)!} x^{l-1} dx \qquad (64)$$

$$= \tfrac{1}{l!} \frac{1}{\left((\text{SNR}')^*\right)^l}$$

$$\approx \frac{1}{\left((\text{SNR}')^*\right)^l},$$

where the term $\tfrac{1}{l!}$ is negligible.

Specifically, from (58) and (64), the $\delta$ *manifold parameter* picks up the following value in the single-carrier CVQKD setting:

$$\delta_{single} = 1, \qquad (65)$$

while in the multicarrier CVQKD setting,

$$\delta_{AMQD} = l. \qquad (66)$$

The result in (66) will be further sharpened in Section 3 since it significantly depends on the properties of the properties of the corresponding phase space constellation $\mathcal{C}_\mathcal{S}(\mathcal{N})$. From (69) it clearly follows that the extractable manifold $\delta$ determines the error probability of the transmission, and for higher $\delta$, the reliability of the transmission improves.

Particularly, in a *multiple-access* CVQKD scenario, there exists another degree of freedom in the channel, the number of information carriers allocated to a given user $U$. This type of degree of freedom is denoted by $\varsigma$ and is referred to as the *degree of freedom ratio*. Without loss of generality, in the function of $\varsigma > 0$ (65) and (66) precisely can be rewritten as

$$\delta_{single} = 1 - \varsigma, \qquad (67)$$

while, in the multicarrier CVQKD setting, it refers to the ratio of the subcarriers allocated to a given user,

$$\delta_{AMQD} = l(1 - \varsigma). \qquad (68)$$

Thus, in a multicarrier CVQKD scenario with $l$ Gaussian sub-channels, for a given $\varsigma > 0$, the overall gain is $l$. As follows, using (67) and (68), the error probabilities can be rewritten precisely as

$$p_{err}^{single} = \frac{1}{\left((\text{SNR}')^*\right)^{\delta_{single}}} = \frac{1}{\left((\text{SNR}')^*\right)^{(1-\varsigma)}}, \qquad (69)$$

$$p_{err}^{AMQD} = \frac{1}{\left((\text{SNR}')^*\right)^{\delta_{AMQD}}} = \frac{1}{\left((\text{SNR}')^*\right)^{l(1-\varsigma)}}. \qquad (70)$$

The $p_{err}^{single}$ and $p_{err}^{AMQD}$ error probabilities of (69) and (70) for $(\text{SNR}')^* \geq 1$, for $l = 5, 10$ Gaussian sub-channels, and at $\varsigma = 0.6$ are compared in Fig. 1.

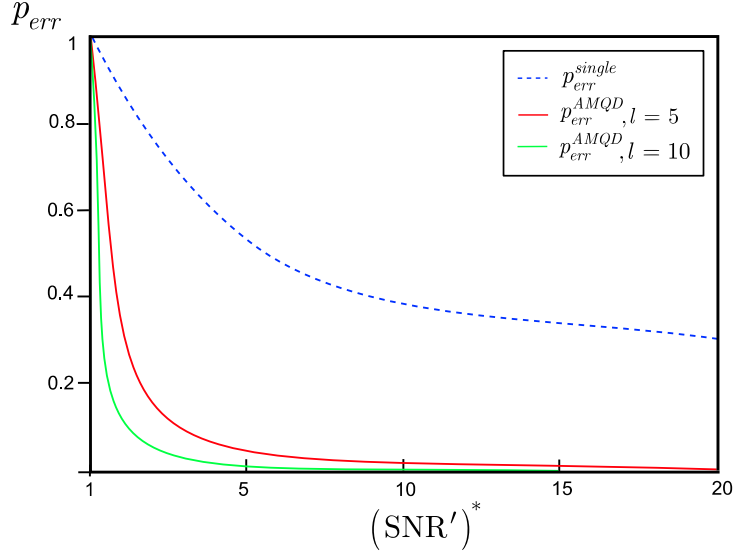

**Figure 1.** The error probabilities in the single-carrier and multicarrier CVQKD, $(\mathrm{SNR}')^* \geq 1$, $l = 5, 10$, and $\varsigma = 0.6$.

In a multicarrier CVQKD protocol run, there exists an optimal tradeoff between $\delta$ and $\varsigma$; however, it requires to make some preliminary assumptions, as it is concluded in Lemma 1.

**Lemma 1.** (Manifold extraction in multicarrier CVQKD). *For any $S'_k(\mathcal{N}) > 0$ and $\varsigma_k = \frac{1}{P'(\mathcal{N})} S'_k(\mathcal{N}) n_{\min}$, $\varsigma_k > 0$ of user $U_k, k = 1,\ldots, K_{out}$, the $\delta_k(\varsigma_k)$ extractable manifold is the ratio of $p_{err}(S'_k(\mathcal{N}))$ error probability and the $n_{\min}$-normalized private classical capacity $\frac{1}{n_{\min}} P'(\mathcal{N})$, derived at the asymptotic limit of $(\mathrm{SNR}')^* \to \infty$.*

*Proof.*
In particular, at a given $\varsigma_k$ and $S'_k(\mathcal{N})$ (see (52)), the $\delta_k(\varsigma_k)$ manifold parameter of user $U_k, k = 1,\ldots, K_{out}$ is as follows:

$$\delta_k(\varsigma_k) = \lim_{(\mathrm{SNR}')^* \to \infty} \frac{-\log_2 p_{err}(S'_k(\mathcal{N}))}{\frac{1}{n_{\min}} P'(\mathcal{N})}, \tag{71}$$

where $p_{err}(S'_k(\mathcal{N}))$ is the error probability of $U_k$ at $S'_k(\mathcal{N})$, while $(\mathrm{SNR}')^*$ is the SNR of $\mathcal{N}$ in an SVD-assisted AMQD modulation for private information transmission (see (44)).
Specifically, assuming that the condition of

$$\tfrac{1}{l} \sum_l |F(T_i(\mathcal{N}_i))| \sqrt{(\mathrm{SNR}')^*} \partial_k \geq c, \tag{72}$$

holds, where $c > 0$ is a constant and $\partial_k$ is the minimum distance of the $2^{S'_k(\mathcal{N})}$ normalized constellation points $|\varphi_i\rangle, |\varphi_j\rangle, j \neq i$ in a phase space constellation $\mathcal{C}'_\mathcal{S}(\mathcal{N})$, $\mathcal{C}'_\mathcal{S}(\mathcal{N}) \subseteq \mathcal{C}_\mathcal{S}(\mathcal{N})$, $|\mathcal{C}'_\mathcal{S}(\mathcal{N})| = 2^{S'_k(\mathcal{N})}$, evaluated precisely as

$$\partial_k = \frac{1}{2^{S'_k(\mathcal{N})/2}}, \tag{73}$$

then at a given secret key rate $S'_k(\mathcal{N})$, the $p_{err}(S'_k(\mathcal{N}))$ error probability of the transmission of $U_k$ decays as

$$\begin{aligned} p_{err}(S'_k(\mathcal{N})) &= Q\left(\sqrt{\tfrac{1}{l}\sum_l |F(T_i(\mathcal{N}_i))|^2 \tfrac{(\mathrm{SNR}')^*}{2} \partial_k^2}\right) \\ &= \frac{2^{S'_k(\mathcal{N})}-1}{(\mathrm{SNR}')^*}, \end{aligned} \tag{74}$$

where $Q(\cdot)$ is the Gaussian tail function. Note that the condition of (72) follows from the fact that the separation (i.e., $\partial_k$) of the constellation points of $\mathcal{C}_\mathcal{S}(\mathcal{N})$ has to be significantly larger than $\sigma_\mathcal{N}$; otherwise, the $Q(\cdot)$ Gaussian tail function yields in high error probabilities [20]. Without loss of generality, for an $\mathcal{N}_i$ dedicated to $U_k$ with $S'_k(\mathcal{N}_i)$, the phase space constellation is referred to as $\mathcal{C}'_\mathcal{S}(\mathcal{N}_i)$, $|\mathcal{C}'_\mathcal{S}(\mathcal{N}_i)| = 2^{S'_k(\mathcal{N}_i)}$, and $\partial_{k,i} = \frac{1}{2^{S'_k(\mathcal{N}_i)/2}}$, and

$$\begin{aligned} p_{err}(S'_k(\mathcal{N}_i)) &= Q\left(\sqrt{|F(T_i(\mathcal{N}_i))|^2 \tfrac{(\mathrm{SNR}'_i)^*}{2} \partial_{k,i}^2}\right) \\ &= \frac{2^{S'_k(\mathcal{N}_i)}-1}{(\mathrm{SNR}'_i)^*}. \end{aligned} \tag{75}$$

Exploiting the argumentation of (61) on the averaging over the $\mathcal{S}$ statistics of the channel transmittance coefficients, and the related result in (59), the $p_{err}(S'_k(\mathcal{N}_i))$ of a given Gaussian sub-channel $\mathcal{N}_i$ at $S'_k(\mathcal{N}_i)$ can be determined precisely as

$$p_{err}(S'_k(\mathcal{N}_i)) = 1 - e^{-\frac{\left(2^{S'_k(\mathcal{N}_i)}-1\right)}{(\mathrm{SNR}'_i)^*}}, \tag{76}$$

which at $(\mathrm{SNR}'_i)^* \to \infty$ coincidences with (75).

Thus, using $P_i \in \mathcal{U}$, $i = 2,...,l$ drawn from a $\mathcal{U}$ uniform distribution, the *private* permutation phase space constellation $\mathcal{C}'^P_\mathcal{S}(\mathcal{N})$, $\mathcal{C}'^P_\mathcal{S}(\mathcal{N}) \subseteq \mathcal{C}^P_\mathcal{S}(\mathcal{N})$, $|\mathcal{C}'^P_\mathcal{S}(\mathcal{N})| = 2^{S'_k(\mathcal{N})}$ can be defined for the private multicarrier transmission as

$$\begin{aligned} \mathcal{C}'^P_\mathcal{S}(\mathcal{N}) &= (\mathcal{C}'_\mathcal{S}(\mathcal{N}_1),...,\mathcal{C}'_\mathcal{S}(\mathcal{N}_l)) \\ &= (\mathcal{C}'_\mathcal{S}(\mathcal{N}_1), P_2 \mathcal{C}'_\mathcal{S}(\mathcal{N}_1),...,P_l \mathcal{C}'_\mathcal{S}(\mathcal{N}_1)), \end{aligned} \tag{77}$$

where $d_{\mathcal{C}'_\mathcal{S}(\mathcal{N}_i)} = d_{\mathcal{C}'_\mathcal{S}(\mathcal{N}_j)}$ is the cardinality of $\mathcal{C}'_\mathcal{S}(\mathcal{N}_i)$. The private permutation phase space constellation of (77) can be used as a corresponding $\mathcal{C}'_\mathcal{S}(\mathcal{N}_i)$, for each $\mathcal{N}_i$ sub-channels.

In particular, assuming the use of $\mathcal{S}(F(T_i(\mathcal{N}_i)))$ in (61), the $p_{err}(S'_k(\mathcal{N}_i))$ of a given sub-channel $\mathcal{N}_i$ with secret key rate $S'_k(\mathcal{N}) = \frac{S_k}{n_{\min}} P'(\mathcal{N})$, can be rewritten precisely as

$$p_{err}\left(S'_k\left(\mathcal{N}_i\right)\right) = \Pr\left(\log_2\left(1 + \left|F\left(T_i\left(\mathcal{N}_i\right)\right)\right|^2 \left(\mathrm{SNR}'_i\right)^*\right) < S'_k\left(\mathcal{N}_i\right)\right)$$

$$= \Pr\left(\left|F\left(T_i\left(\mathcal{N}_i\right)\right)\right|^2 < \frac{\left(\left(\mathrm{SNR}'_i\right)^*\right)^{\varsigma_{k,i}} - 1}{\left(\mathrm{SNR}'_i\right)^*}\right) \quad (78)$$

$$\approx \frac{1}{\left(\left(\mathrm{SNR}'_i\right)^*\right)^{1-\varsigma_{k,i}}} = \frac{1}{\left(\left(\mathrm{SNR}'_i\right)^*\right)^{\delta_{k,i}\left(\varsigma_{k,i}\right)}},$$

where $\Pr\left(\left|F\left(T_i\left(\mathcal{N}_i\right)\right)\right|^2 < x\right) \approx x$, by theory at the distribution of (61) [20–22], and $\delta_{k,i}\left(\varsigma_{k,i}\right) = 1 - \varsigma_{k,i}$. Specifically, if $F\left(T_i\left(\mathcal{N}_i\right)\right) \in \mathcal{CN}\left(0, \sigma^2_{F\left(T_i(\mathcal{N}_i)\right)}\right)$ for all $\mathcal{N}_i$, then for the $S'(\mathcal{N})$ secret key rate in the low SNR regimes the following result yields, precisely:

$$S'(\mathcal{N}) \leq \mathbb{E}\left[\log_2\left(1 + \frac{\sigma^2_{\omega''} \max_i\left|F\left(T_i\left(\mathcal{N}_i\right)\right)\right|^2}{\sigma^2_{\mathcal{N}^*}}\right)\right]$$

$$\approx \mathbb{E}\left(\frac{\sigma^2_{\omega''} \max_i\left|F\left(T_i\left(\mathcal{N}_i\right)\right)\right|^2}{\sigma^2_{\mathcal{N}^*}}\right)\log_2 e \approx \frac{\sigma^2_{\omega''}}{\sigma^2_{\mathcal{N}^*}}\log_2 e, \quad (79)$$

while in the high SNR regimes

$$S'(\mathcal{N}) \leq \mathbb{E}\left[\log_2\left(\frac{\sigma^2_{\omega''} \max_i\left|F\left(T_i\left(\mathcal{N}_i\right)\right)\right|^2}{\sigma^2_{\mathcal{N}^*}}\right)\right]$$

$$\approx \log_2 \frac{\sigma^2_{\omega''}}{\sigma^2_{\mathcal{N}^*}} + \mathbb{E}\left(\log_2\left(\max_i\left|F\left(T_i\left(\mathcal{N}_i\right)\right)\right|^2\right)\right), \quad (80)$$

and from the law of large numbers [20]:

$$\frac{1}{l}\left(\sum_l \log_2\left(1 + \frac{\sigma^2_{\omega''_i}\left|F\left(T_i\left(\mathcal{N}_i\right)\right)\right|^2}{\sigma^2_{\mathcal{N}^*_i}}\right)\right) = \mathbb{E}\left[\log_2\left(1 + \frac{\sigma^2_{\omega''} \max_i\left|F\left(T_i\left(\mathcal{N}_i\right)\right)\right|^2}{\sigma^2_{\mathcal{N}^*}}\right)\right]. \quad (81)$$

At $F\left(T_i\left(\mathcal{N}_i\right)\right) \in \mathcal{CN}\left(0, \sigma^2_{F\left(T_i(\mathcal{N}_i)\right)}\right)$, the density of (76) is depicted in Fig. 2.

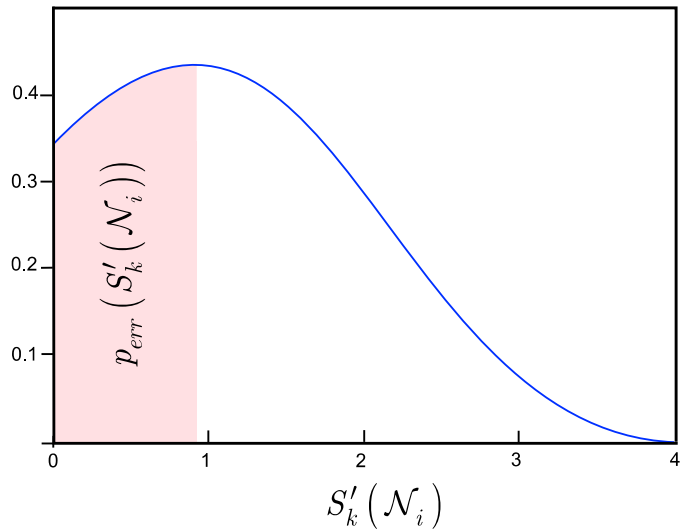

**Figure 2.** At the distribution of the $\mathcal{S}$ statistics of (61), the shaded area under the curve gives the $p_{err}$ error probability at a given $S'_k\left(\mathcal{N}_i\right)$ for a Gaussian sub-channel $\mathcal{N}_i$.

Note that for the transmission of classical (i.e., non-private) information, $\mathcal{C}_\mathcal{S}$ has a cardinality of $|\mathcal{C}_\mathcal{S}| = 2^{R'_k}$ at a given $R'_k(\mathcal{N})$, with a corresponding $\text{SNR}'$ and $\partial_k = \frac{1}{2^{R'_k(\mathcal{N})/2}}$; for the details, see the description of [4].

An error event $\text{E}_{err}$ of (54) for a sub-channel $\mathcal{N}_i$ can be rewritten as

$$\text{E}_{err} : \tfrac{1}{l}\sum_l |F(T_i(\mathcal{N}_i))|^2 < \frac{2^{S'_k(\mathcal{N})}-1}{(\text{SNR}')^*}; \qquad (82)$$

thus, introducing $\text{Z} > 1$ which brings up by the use of a corresponding $\mathcal{C}_\mathcal{S}$ (such as the permutation scheme in [4]), a typical error probability is precisely expressed as follows:

$$\begin{aligned}
p_{err}(S'_k(\mathcal{N}_i)) &= \Pr\left(\tfrac{1}{l}\sum_l \log_2\left(1 + |F(T_i(\mathcal{N}_i))|^2 (\text{SNR}'_i)^*\right) < S'_k(\mathcal{N}_i)\right) \\
&= \Pr\left(\tfrac{1}{l}\sum_l \log_2\left(1 + |F(T_i(\mathcal{N}_i))|^2 (\text{SNR}'_i)^*\right) < \tfrac{\varsigma_{k,i}}{n_{\min}} P'(\mathcal{N}_i)\right) \\
&= \Pr\left(\tfrac{1}{l}\sum_l |F(T_i(\mathcal{N}_i))|^2 < \frac{\left((\text{SNR}')^*\right)^{\varsigma_k}-1}{(\text{SNR}')^*}\right) \\
&= \frac{1}{\left((\text{SNR}')^*\right)^{1-\varsigma_k}} = \frac{1}{\left((\text{SNR}')^*\right)^{\text{Z}(1-\varsigma_k)}} \\
&= \left(\frac{\left((\text{SNR}')^*\right)^{\varsigma_k}-1}{(\text{SNR}')^*}\right)^{\text{Z}},
\end{aligned} \qquad (83)$$

where $\Pr\left(|F(T_i(\mathcal{N}_i))|^2 < x\right) \approx x^{\text{Z}}$ as $\left(\left((\text{SNR}')^*\right)^{\varsigma_k}-1\right)/(\text{SNR}')^* \to 0$, by theory. (Note that in (83), the $T_i$ transmittance coefficients are arbitrarily distributed, in contrast to (78)). As one immediately can conclude, the phase space constellation $\mathcal{C}_\mathcal{S}$ provides a further decreased $p_{err}$ in comparison to (70).

Putting the pieces together, the optimal manifold-degree of freedom ratio *tradeoff curve* [20] $f$ for a single-carrier scheme (e.g., if $l = 1$ we trivially have a single-carrier scheme) can be expressed as

$$f(\mathcal{N}_i) : \delta_k(\varsigma_k) = \text{Z}(1-\varsigma_k), \qquad (84)$$

where $0 < \varsigma_k \leq 1$.

Specifically, some calculations then straightforwardly reveal that any phase space constellation $\mathcal{C}_\mathcal{S}(\mathcal{N})$ that satisfies the condition of

$$\partial_k^2 = q\frac{1}{2^{S'_k(\mathcal{N})}} \qquad (85)$$

achieves the optimal $f$ tradeoff curve, for any constant $q > 0$. The recently proposed permutation phase space constellation $\mathcal{C}_\mathcal{S}^P(\mathcal{N})$ for SVD-assisted AMQD provably satisfies this condition [4].

Without loss of generality, assuming a constant $g > 0$, $p_{err}$ can be rewritten precisely as

$$p_{err}^g\left(S_k'\left(\mathcal{N}_i\right)\right) : \Pr\left[\tfrac{1}{l}\sum_l \log_2\left(1+\left|F\left(T_i\left(\mathcal{N}_i\right)\right)\right|^2\left(\left(\mathrm{SNR}_i'\right)^*\right)^{1-g}\right) < S_k'\left(\mathcal{N}_i\right)\right]$$

$$= \Pr\left[\tfrac{1}{l}\sum_l \left|F\left(T_i\left(\mathcal{N}_i\right)\right)\right|^2 < \frac{\left[\left(\left(\mathrm{SNR}'\right)^*\right)^{1-g}\right]^{\varsigma_k}-1}{\left(\left(\mathrm{SNR}'\right)^*\right)^{1-g}}\right] \quad (86)$$

$$= \frac{\left(\left(\mathrm{SNR}'\right)^*\right)^{\varsigma_k(1-g)}-1}{\left(\left(\mathrm{SNR}'\right)^*\right)^{1-g}}.$$

Thus, for $\mathrm{E}_{err}^g$, the manifold parameter is $\delta_k^g\left(\varsigma_k\right)$ as

$$\delta_k^g\left(\varsigma_k\right) = \delta_k\left(\varsigma_k\right)\cdot\left(1-g\right) \geq \lim_{\left(\mathrm{SNR}'\right)^*\to\infty} \frac{-\log_2 p_{err}\left(S_k'(\mathcal{N})\right)}{\tfrac{1}{n_{\min}}P'(\mathcal{N})}. \quad (87)$$

∎

In Section 3, we give a proof on the multicarrier CVQKD scenario and show that there exists an optimal tradeoff between $\varsigma_k$ and $\delta_k$ for any $U_k$. We further reveal that in a multicarrier setting, the manifold extraction significantly exceeds the possibilities of a single-carrier CVQKD scenario.

## 3 Multidimensional Manifold Space

**Theorem 1** (Multidimensional manifold space for multicarrier CVQKD). *For any $K_{in}, K_{out}$, and $\varsigma_k > 0$, the $\mathcal{M}$ manifold is a $\dim(\mathcal{M}) = K_{in}\varsigma_k + \left(K_{out}-\varsigma_k\right)\varsigma_k$ dimensional space. The number $N_{\dim^\perp}$ of dimensions orthogonal to $\mathcal{M}$ in the $\dim\left(\mathrm{S}\left(F\left(\mathbf{T}(\mathcal{N})\right)\right)\right) = K_{in}K_{out}$ dimensional space $\mathrm{S}\left(F\left(\mathbf{T}(\mathcal{N})\right)\right)$ is $N_{\dim^\perp} = K_{in}K_{out} - \left(K_{in}\varsigma_k + \left(K_{out}-\varsigma_k\right)\varsigma_k\right) = \left(K_{in}-\varsigma_k\right)\left(K_{out}-\varsigma_k\right)$, with $p_{err}\left(\varsigma_k\right) = \frac{1}{\left(\left(\mathrm{SNR}'\right)^*\right)^{N_{\dim^\perp}}}$, and optimal tradeoff curve $h(\mathcal{M}) : \delta_k\left(\varsigma_k\right) = N_{\dim^\perp}$.*

*Proof.*
The proof assumes a $K_{in}, K_{out}$ multiuser scenario. First we express $p_{err}$ as follows:

$$p_{err} = \Pr\left(\mathrm{E}_{err} = \sum_{i=1}^{n_{\min}}\log_2\left(1+\tfrac{(\mathrm{SNR}')^*}{K_{in}}\lambda_i^2\right) < \tfrac{\varsigma_k(r+K_{in}-1)}{r}P'(\mathcal{N})\right), \quad (88)$$

where $\lambda_i^2$ are the squared random singular values of $F\left(\mathbf{T}(\mathcal{N})\right)$ [20–22].

Assuming $\varsigma_k > 0$, the $\lambda_i$ singular values can be decomposed into subsets $s_0$ and $s_1$ such that set

$$s_0 = \left\{\lambda_1, \ldots, \lambda_{\varsigma_k}\right\} \quad (89)$$

contains the largest $\varsigma_k$ singular values of $F\left(\mathbf{T}(\mathcal{N})\right)$, $\lambda_i < \lambda_{i+1}$, $i = 1, \ldots, \varsigma_k$, and where

$$\max_{\forall i}\left\{\lambda_1, \ldots, \lambda_{\varsigma_k}\right\} \leq 1. \quad (90)$$

The remaining $n_{\min} - \varsigma_k$ singular values formulate the subset $s_1$, as

$$s_1 = \left\{\lambda_{\varsigma_k+1}, \ldots, \lambda_{n_{\min}}\right\}, \quad (91)$$

where
$$\max_{\forall i}\{\lambda_{\varsigma_k+1},\ldots,\lambda_{n_{\min}}\} \leq \frac{1}{(\text{SNR}')^*}. \tag{92}$$

In particular, from (89) and (91), for the rank of $F(\mathbf{T}(\mathcal{N}))$, the following relation identifies an error event $\mathrm{E}_{err}$:
$$\mathrm{E}_{err}: rank(F(\mathbf{T}(\mathcal{N}))) \leq \varsigma_k. \tag{93}$$

Thus, the $p_{err}$ at a given $\varsigma_k$ is precisely referred to as
$$p_{err}(\varsigma_k) = \Pr(rank(F(\mathbf{T}(\mathcal{N}))) \leq \varsigma_k). \tag{94}$$

Specifically, at $\varsigma_k = 0$, $p_{err}(\varsigma_k)$ is evaluated as
$$p_{err}(\varsigma_k = 0) = \Pr\left(F(\mathbf{T}(\mathcal{N})): \max_{\forall i}\{\lambda_1,\ldots,\lambda_{n_{\min}}\} \leq \frac{1}{(\text{SNR}')^*}\right)$$
$$= \frac{1}{\left((\text{SNR}')^*\right)^{K_{in}K_{out}}}. \tag{95}$$

At $\varsigma_k \to 0$, in (88) the corresponding relation is
$$\lambda_i \leq \frac{1}{(\text{SNR}')^*}, \tag{96}$$

thus for the sum of the $n_{\min}$ squared eigenvalues $\lambda_i$,
$$\sum_{i=1}^{n_{\min}} \lambda_i^2 = \sum_k \left|F(T(\mathcal{N}_{U_k}))\right|^2, \tag{97}$$

where $\mathcal{N}_{U_k}$ refers to the logical channel of $U_k$, which consists of the allocated Gaussian subcarriers of that user.

As follows, (96) holds if only
$$\left|F(T(\mathcal{N}_{U_k}))\right|^2 \leq \frac{1}{(\text{SNR}')^*}, \tag{98}$$

thus,
$$p_{err}(\varsigma_k \to 0) = \Pr\left(\left|F(T(\mathcal{N}_{U_k}))\right|^2 < \frac{1}{(\text{SNR}')^*}\right)$$
$$\approx \frac{1}{\left((\text{SNR}')^*\right)^{K_{in}K_{out}}} \tag{99}$$
$$= \frac{1}{\left((\text{SNR}')^*\right)^{\dim(F(\mathbf{T}(\mathcal{N})))}},$$

since the dimension of the space $\mathrm{S}(F(\mathbf{T}(\mathcal{N})))$ for a $K_{in}, K_{out}$ multiple-access scenario is [20]:
$$\dim(\mathrm{S}(F(\mathbf{T}(\mathcal{N})))) = K_{in}K_{out}. \tag{100}$$

Particularly, the $\mathcal{M}$ manifold space can be represented as a $rank-\varsigma_k$ matrix $M$ (see later (102)); thus, it precisely has a dimension of
$$\dim(\mathcal{M}) = K_{in}\varsigma_k + (K_{out} - \varsigma_k)\varsigma_k. \tag{101}$$

Specifically, for any $\varsigma_k > 0$,

$$p_{err}(\varsigma_k > 0) = \Pr\left(rank\left(F\left(\mathbf{T}(\mathcal{N})\right)\right) \leq rank(M) = \varsigma_k\right)$$
$$= \frac{1}{\left((\mathrm{SNR}')^*\right)^{N_{\dim^\perp}}}, \qquad (102)$$

where $N_{\dim^\perp}$ refers to the $\dim^\perp$ dimensions orthogonal to $\mathcal{M}$ in the $\dim\left(\mathrm{S}\left(F\left(\mathbf{T}(\mathcal{N})\right)\right)\right) = K_{in}K_{out}$ dimensional space S of $F\left(\mathbf{T}(\mathcal{N})\right)$, and $M$ is a matrix with rank $\varsigma_k$.

Thus, for $\varsigma_k > 0$, $p_{err}(\varsigma_k)$ is related to as the difference between matrix $F\left(\mathbf{T}(\mathcal{N})\right)$ and a rank-$\varsigma_k$ matrix $M$, since $F\left(\mathbf{T}(\mathcal{N})\right)$ is a $K_{in} \times K_{out}$ matrix with $rank = \varsigma_k$; that is, it has $\varsigma_k > 0$ linearly independent row vectors from the $K_{out}$ rows [20–22]. Without loss of generality, $M$ can be characterized by $K_{in}\varsigma_k + (K_{out} - \varsigma_k)\varsigma_k$ parameters by theory, from which (101) straightforwardly follows.

Putting the pieces together, the $N_{\dim^\perp}$ number of $\dim^\perp$ dimensions orthogonal to manifold space $\mathcal{M}$ in the $\dim\left(\mathrm{S}\left(F\left(\mathbf{T}(\mathcal{N})\right)\right)\right)$ dimensional space of $F\left(\mathbf{T}(\mathcal{N})\right)$ is precisely

$$\begin{aligned}N_{\dim^\perp} &= K_{in}K_{out} - \dim(\mathcal{M}) \\ &= K_{in}K_{out} - \left(K_{in}\varsigma_k + (K_{out} - \varsigma_k)\varsigma_k\right) \\ &= (K_{in} - \varsigma_k)(K_{out} - \varsigma_k).\end{aligned} \qquad (103)$$

Particularly, from (102) the multidimensional optimal tradeoff function is yielded as
$$h(\mathcal{M}): \delta(\varsigma_k) = N_{\dim^\perp}. \qquad (104)$$

The $N_{\dim^\perp}$ in function of $\dim(\mathcal{M})$, at $K_{in} = K_{out} - 1$, $\varsigma_k = 0,3;0.6;0.9$ is depicted in Fig. 3.

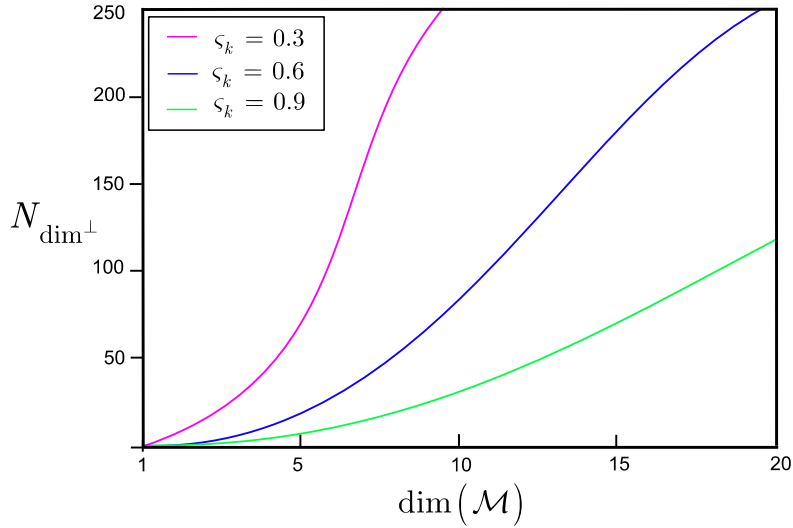

**Figure 3.** The values of $N_{\dim^\perp}$ in function of $\dim(\mathcal{M})$, at $K_{in} = K_{out} - 1$, $\varsigma_k = 0,3;0.6;0.9$.

■

# 4 Manifold Extraction for Multicarrier CVQKD

**Theorem 2** (Manifold extraction for multicarrier CVQKD). *For user $U_k$, the manifold extraction at $l$ Gaussian sub-channels leads to an optimal $h$ tradeoff curve $h: \delta_k(\varsigma_k) = l\mathrm{Z}(1-\varsigma_k) = lf$ for any multicarrier scheme, where $f$ is the optimal tradeoff curve of a single-carrier CVQKD protocol, $f: \delta_k(\varsigma_k) = \mathrm{Z}(1-\varsigma_k)$, and $\mathrm{Z} > 1$.*

*Proof.*
In the first part of the proof, we assume the case $l = 1$, which is analogous to a single-carrier transmission. In the second part of the proof, we study the multicarrier case for $l$ Gaussian sub-channels and reveal that a multicarrier case allows significantly improved manifold extraction.

In the single-carrier scenario, the phase space constellation $\mathcal{C}'_\mathcal{S}(\mathcal{N})$, $\mathcal{C}'_\mathcal{S}(\mathcal{N}) \subseteq \mathcal{C}_\mathcal{S}(\mathcal{N})$, $|\mathcal{C}'_\mathcal{S}(\mathcal{N})| = 2^{S'_k(\mathcal{N})}$, leads to $\partial_k = \frac{1}{2^{S'_k(\mathcal{N})/2}}$, $p_{err}(S'_k(\mathcal{N})) = \frac{2^{S'_k(\mathcal{N})}-1}{(\mathrm{SNR}')^*}$, and $f : \delta_k(\varsigma_k) = \mathrm{Z}(1-\varsigma_k)$, as it has been already shown in the proof of Lemma 1. This is precisely the situation for $l = 1$.

Specifically, for the multicarrier CVQKD case, let us assume that there are $l$ Gaussian sub-channels dedicated to the transmission of the $d_i, i = 1,\ldots,l$ Gaussian subcarriers, with $(\mathrm{SNR}'_i)^* = \frac{\sigma^2_{\omega''_i}}{\sigma^2_{\mathcal{N}^*_i}}$ per $\mathcal{N}_i$, and secret key rate $S'_k(\mathcal{N}_i)$ as

$$S'_k(\mathcal{N}_i) = \frac{\varsigma_{k,i}}{n_{\min}} P'(\mathcal{N}_i). \qquad (105)$$

Without loss of generality, for the total constraint of SVD-assisted AMQD [4], one has precisely

$$\sum_{i=1}^{l} \log_2\left(1 + |F(T_i(\mathcal{N}_i))|^2 (\mathrm{SNR}'_i)^*\right) \geq lS'_k(\mathcal{N}_i). \qquad (106)$$

In particular, by further exploiting the results of SVD-assisted AMQD [4] and following the derivations of Section 5 in [4], here we determine the private random codeword difference for two $l$-dimensional input codewords $\mathbf{p}_A = (p_{A,1},\ldots p_{A,l})^T$ and $\mathbf{p}_B = (p_{B,1},\ldots p_{B,l})^T$. The probability that $\mathbf{p}_A$ is distorted onto $\mathbf{p}_B$ conditioned on $F(\mathbf{T}(\mathcal{N}))$ is evaluated precisely as follows:

$$\Pr(\mathbf{p}_A \to \mathbf{p}_B | F(\mathbf{T}(\mathcal{N}))) = Q\left(\sqrt{\frac{\sigma^2_{\omega''}}{2\sigma^2_{\mathcal{N}^*}} \sum_l |F(T_i(\mathcal{N}_i))|^2 |\partial_i|^2}\right), \qquad (107)$$

where $\partial_i$ is the normalized difference of $p_{A,i}$ and $p_{B,i}$, calculated as follows:

$$\partial_i = \frac{1}{\sqrt{\frac{\sigma^2_{\omega''_i}}{\sigma^2_{\mathcal{N}^*}}}} (p_{A,i} - p_{B,i}). \qquad (108)$$

Assuming the case that in (107), the condition

$$\frac{\sigma^2_{\omega''}}{2\sigma^2_{\mathcal{N}^*}} \sum_l |F(T_i(\mathcal{N}_i))|^2 |\partial_i|^2 < 1 \qquad (109)$$

holds, one obtains

$$\left|\partial_{1\ldots l}\right|^2 > c^l \frac{1}{l^l 2^{S'_k(\mathcal{N}_i)l}}, \tag{110}$$

for any constant $c > 0$ and for an arbitrary pair of $\mathbf{p}_A$ and $\mathbf{p}_B$ [4].

In particular, at a secret key rate $S'_k$ per $\mathcal{N}_i$, the cardinality of $\mathcal{C}'_\mathcal{S}(\mathcal{N}_i)$ is as follows:

$$\left|\mathcal{C}'_\mathcal{S}(\mathcal{N}_i)\right| = 2^{S'_k(\mathcal{N}_i)}. \tag{111}$$

Thus, in the private transmission each $\mathcal{C}'_\mathcal{S}(\mathcal{N}_i)$ is precisely defined with $2^{S'_k(\mathcal{N}_i)}$ CV states $|\phi_i\rangle$ for each $\mathcal{N}_i$, at an averaged $\tilde{S}'_k(\mathcal{N}_i)$ as

$$\left|\mathcal{C}'_\mathcal{S}(\mathcal{N})\right| = 2^{\sum_i S'_k(\mathcal{N}_i)} \approx 2^{l\tilde{S}'_k(\mathcal{N}_i)}. \tag{112}$$

Specifically, evaluating the $Q(\cdot)$ Gaussian tail function at

$$\min_{\forall F(T_i(\mathcal{N}_i))} |F(T_i(\mathcal{N}_i))|^2 = \min_{\forall i} \frac{1}{\frac{\sigma^2_{\omega''}}{\sigma^2_{\mathcal{N}^*}}} \left( v_{Eve} \frac{1}{|\partial_i|^2} - 1 \right), \tag{113}$$

where $v_{Eve}$ is Eve's corresponding security parameter in an optimal Gaussian attack (for an exact derivation of this parameter, see the description of the AMQD modulation in [2]).

The result in (113) leads to a worst-case scenario precisely as

$$\tilde{p}_{err} = \Pr\left(\mathbf{p}_A \to \mathbf{p}_B \middle| F(\mathbf{T}(\mathcal{N}))\right) = Q\left(\sqrt{\min_{\forall F(T_i(\mathcal{N}_i))} \frac{\sigma^2_{\omega''}}{2\sigma^2_{\mathcal{N}^*}} \sum_l |F(T_i(\mathcal{N}_i))|^2 |\partial_i|^2}\right), \tag{114}$$

such that for the $l$ $\mathcal{N}_i$ Gaussian sub-channels, the following constraint is satisfied:

$$\log_2\left(1 + \frac{\sigma^2_{\omega''}|F(\mathbf{T}(\mathcal{N}))|^2}{\sigma^2_{\mathcal{N}^*}}\right) = \sum_l \log_2\left(1 + \frac{\sigma^2_{\omega''}|F(T_i(\mathcal{N}_i))|^2}{\sigma^2_{\mathcal{N}^*}}\right) \geq l\tilde{S}'_k(\mathcal{N}_i). \tag{115}$$

In particular, the optimal manifold extraction $\delta_k(\varsigma_k)$ requires the maximization of the product distance $|\partial_{1\ldots l}|^{2/l}$ at (114); thus without loss of generality, the optimizing condition at $l$ Gaussian sub-channels is a maximization as:

$$\delta_k(\varsigma_k) : \max_{\forall i} |\partial_{1\ldots l}|^{2/l} > c \frac{1}{l 2^{S'_k(\mathcal{N}_i)}}. \tag{116}$$

Since for $\mathcal{C}^P_\mathcal{S}(\mathcal{N}_i)$ this condition is satisfied, by using the $\mathcal{C}^P_\mathcal{S}(\mathcal{N}_i)$ random permutation operators defined in [4] as $\mathcal{C}_\mathcal{S}(\mathcal{N}_i)$ for the Gaussian sub-channels, the optimality of $\delta_k(\varsigma_k)$ can be satisfied. Using (108), the constraint of (115) can be rewritten as follows [4], [20]:

$$\sum_l \log_2\left(1 + \frac{\tilde{Q}_i(\mathcal{N}_i)}{|\partial_i|^2}\right) \geq l\tilde{S}'_k(\mathcal{N}_i), \tag{117}$$

where

$$\tilde{Q}_i(\mathcal{N}_i) = \frac{|F(T_i(\mathcal{N}_i))|^2 |\partial_i|^2 \sigma^2_{\omega''}}{\sigma^2_{\mathcal{N}^*}} \tag{118}$$

and

$$\min_{\forall \tilde{Q}_i \geq 0} \frac{1}{2} \sum_l \tilde{Q}_i = \min_{\forall F(T_i(\mathcal{N}_i))} \frac{\sigma^2_{\omega''}}{2\sigma^2_{\mathcal{N}^*}} \sum_l |F(T_i(\mathcal{N}_i))|^2 |\partial_i|^2, \tag{119}$$

where

$$\min_{\forall F(T_i(\mathcal{N}_i))} \sum_l |F(T_i(\mathcal{N}_i))|^2 = \sum_l \frac{1}{\sqrt{\frac{\sigma^2_{\omega''}}{\sigma^2_{\mathcal{N}^*}}}} \left( \nu_{Eve} \frac{1}{|\partial_i|^2} - 1 \right). \tag{120}$$

Without loss of generality, from (119) and (120), the Gaussian tail function in (114) can be precisely rewritten as

$$Q\left( \sqrt{\frac{1}{2} \sum_l \left( \nu_{Eve} - |\partial_i|^2 \right)} \right) \tag{121}$$

and

$$\sum_l \log_2 \left( \nu_{Eve} \frac{1}{|\partial_i|^2} \right) = \sum_l \log_2 \left( 1 + \frac{\tilde{Q}_i(\mathcal{N}_i)}{|\partial_i|^2} \right) = l\tilde{S}'_k(\mathcal{N}_i). \tag{122}$$

Particularly, from these derivations, the manifold extraction for the multicarrier scenario is yielded as follows. The $E_{err}$ error event can be rewritten as

$$E_{err} : \sum_{i=1}^{l} \log_2 \left( 1 + |F(T_i(\mathcal{N}_i))|^2 (SNR'_i)^* \right) < l\tilde{S}'_k(\mathcal{N}_i); \tag{123}$$

thus for $p_{err}(\tilde{S}'_k(\mathcal{N}))$,

$$\begin{aligned} p_{err}(\tilde{S}'_k(\mathcal{N})) &= \Pr\left[ \sum_{i=1}^{l} \log_2 \left( 1 + |F(T_i(\mathcal{N}_i))|^2 (SNR'_i)^* \right) < l\tilde{S}'_k(\mathcal{N}_i) \right] \\ &= \Pr\left[ \sum_{i=1}^{l} \log_2 \left( 1 + |F(T_i(\mathcal{N}_i))|^2 (SNR'_i)^* \right) < l \frac{\varsigma_{k,i}}{n_{\min}} P'(\mathcal{N}_i) \right]. \end{aligned} \tag{124}$$

Specifically, it can be further evaluated as

$$\begin{aligned} p_{err}(\tilde{S}'_k(\mathcal{N})) &= \left( \Pr\left( \log_2 \left( 1 + |F(T_i(\mathcal{N}_i))|^2 (SNR'_i)^* \right) < \frac{\varsigma_{k,i}}{n_{\min}} P'(\mathcal{N}_i) \right) \right)^l \\ &= \left( \Pr\left( |F(T_i(\mathcal{N}_i))|^2 < \frac{((SNR'_i)^*)^{\varsigma_{k,i}} - 1}{(SNR'_i)^*} \right) \right)^l \\ &= \frac{1}{((SNR')^*)^{l(1-\varsigma_{k,i})}} = \left[ \frac{((SNR'_i)^*)^{\varsigma_{k,i}} - 1}{(SNR'_i)^*} \right]^{lZ}, \end{aligned} \tag{125}$$

and at $\varsigma_{k,i}$, the optimal manifold extraction for each $\mathcal{N}_i$ is

$$\delta_{k,i}(\varsigma_{k,i}) = Z(1 - \varsigma_{k,i}) = Z(1 - \varsigma_k). \tag{126}$$

Thus, without loss of generality,

$$p_{err} = \frac{1}{((SNR')^*)^{lZ(1-\varsigma_k)}}. \tag{127}$$

As follows, from (127), the manifold extraction for the $l$ Gaussian sub-channels, and the optimal manifold-degree of freedom ratio tradeoff curve $h$ [20–22] for the multicarrier transmission, is precisely expressed as

$$h : \delta_k(\varsigma_k) = lZ(1 - \varsigma_k) = lf, \tag{128}$$

where $0 \leq \varsigma_k$.

The single-carrier and multicarrier tradeoff curves $f$ and $h$ are compared in Fig. 4.

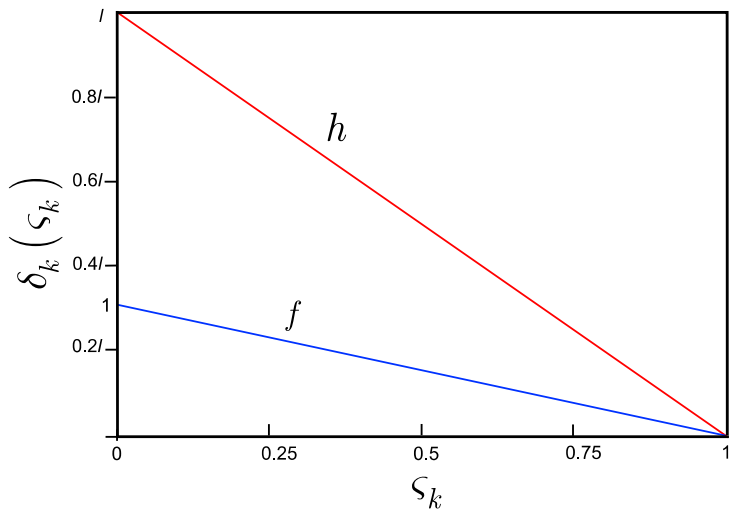

**Figure 4.** Comparison of the optimal manifold-degree of freedom ratio tradeoff curves $f$ and $h$ for single-carrier and multicarrier CVQKD for a user $U_k$.

To conclude the results, the multicarrier CVQKD with $l$ Gaussian sub-channels provides an $l$-fold manifold gain over the single-carrier CVQKD protocols, for all $\varsigma_k$.

∎

## 4.1 Manifold Extraction for AMQD-MQA

**Theorem 3** (Manifold extraction in a multiple-access multicarrier CVQKD). *For any $K_{in}, K_{out}$ multiple-access multicarrier CVQKD scenario, the manifold extraction is maximized via $\delta_k(\varsigma_k) : \max_{\forall i} \left| \lambda_{1\ldots n_{\min}} \right|^{2/n_{\min}}$, where $\lambda_i$ is the i-th smallest singular value of matrix $\mathbf{M}_j = \frac{1}{\sqrt{(\mathrm{SNR}')^*}} (\mathbf{p}_A - \mathbf{p}_B)$, and $\mathbf{p}_A, \mathbf{p}_B$ are l-dimensional random private codewords.*

*Proof.*

Let us assume that $\mathbf{p}_A, \mathbf{p}_B$ are $l$-dimensional inputs, $\mathbf{p}_A = (p_{A,1}, \ldots p_{A,l})^T$ and $\mathbf{p}_B = (p_{B,1}, \ldots p_{B,l})^T$. The $\Pr(\mathbf{p}_A \to \mathbf{p}_B | F(\mathbf{T}(\mathcal{N})))$ pairwise error probability can be evaluated as

$$\Pr(\mathbf{p}_A \to \mathbf{p}_B | F(\mathbf{T}(\mathcal{N}))) = \max_{\forall \eta} Q\left( \frac{F(\mathbf{T}(\mathcal{N})) \cdot (\mathbf{p}_A - \mathbf{p}_B|)}{\sqrt{2}} \right), \quad (129)$$

where $\eta$ is expressed as

$$\eta = F(\mathbf{T}(\mathcal{N})) : \left| F(\mathbf{T}(\mathcal{N})) \right|^2 > \frac{K_{in}\left(2^{S'_k(\mathcal{N})} - 1\right)}{(\mathrm{SNR}')^*}. \quad (130)$$

Without loss of generality, let $\tilde{\lambda}$ be the *smallest eigenvalue* of $\mathbf{M}_j$,

$$\tilde{\lambda} = \min_{\forall i}(\lambda_i), \tag{131}$$

where $\mathbf{M}_j$ stands for the *private codeword difference matrix*,

$$\mathbf{M}_j = \frac{1}{\sqrt{(\mathrm{SNR}')^*}}(\mathbf{p}_A - \mathbf{p}_B). \tag{132}$$

In particular, using (131) and (129) can be written precisely as

$$\Pr\big(\mathbf{p}_A \to \mathbf{p}_B \big| F(\mathbf{T}(\mathcal{N}))\big) = Q\left(\tfrac{1}{2}\tilde{\lambda}^2 K_{in}\left(2^{S'_k(\mathcal{N})} - 1\right)\right). \tag{133}$$

Precisely, the result of (133) follows from the fact that for a $K_{in} \times K_{out}$ matrix $\mathbf{M}_j$, the following relation holds for $\mathbf{M}_j$ and its smallest eigenvalue $\tilde{\lambda}$, by theory:

$$\tilde{\lambda}^2 = \min_{\forall |F(\mathbf{T}(\mathcal{N}))|} F(\mathbf{T}(\mathcal{N}))^\dagger \mathbf{M}_j \mathbf{M}_j^\dagger F(\mathbf{T}(\mathcal{N})). \tag{134}$$

Some calculations then straightforwardly reveal that for $Q\left(\tfrac{1}{2}\tilde{\lambda}^2 K_{in}\left(2^{S'_k(\mathcal{N})} - 1\right)\right) > 1$, the condition on $\tilde{\lambda}$ is as follows:

$$\tilde{\lambda}^2 > \frac{1}{K_{in}\left(2^{S'_k(\mathcal{N})} - 1\right)} \simeq \frac{1}{K_{in} 2^{S'_k(\mathcal{N})}}. \tag{135}$$

Introducing a covariance matrix $\mathbf{K}_o$ as

$$\mathbf{K}_o = (\mathrm{SNR}')^* \frac{I_{K_{in}}}{K_{in}}, \tag{136}$$

where $I_{K_{in}}$ is the $K_{in} \times K_{in}$ identity matrix, $\mathrm{E}_{err}$ can be rewritten as

$$\mathrm{E}_{err} = \log_2 \det\!\left(I_{K_{out}} + F(\mathbf{T}(\mathcal{N}))\mathbf{K}_o F(\mathbf{T}(\mathcal{N}))^\dagger\right) < S'(\mathcal{N}), \tag{137}$$

where without loss of generality,

$$S'(\mathcal{N}) \leq P'(\mathcal{N}) = \max_{\forall i} \mathbb{E}\!\left(\sum_l \log_2\!\left(1 + \frac{\sigma^2_{\omega''_i}|F(T_i(\mathcal{N}_i))|^2}{\sigma^2_{\mathcal{N}^*_i}}\right)\right). \tag{138}$$

Let the SNIR (signal-to noise plus interference ratio) of $\mathcal{N}_i$ be $(\mathrm{SNIR}'_i)^*$ in an SVD-assisted AMQD setting, then $\mathrm{E}_{err}$ can be rewritten precisely as

$$\begin{aligned}\mathrm{E}_{err} &= \log_2 \det\!\left(I_{K_{out}} + F(\mathbf{T}(\mathcal{N}))\frac{(\mathrm{SNR}')^*}{K_{in}} F(\mathbf{T}(\mathcal{N}))^\dagger\right) \\ &= \sum_{i=1}^{K_{in}}\left(1 + (\mathrm{SNIR}'_i)^*\right) < S'(\mathcal{N}).\end{aligned} \tag{139}$$

Then, let us assume that $r$ sub-channels are interfering with each other in the SVD-assisted multicarrier transmission. Specifically, at $r$ interfering sub-channels, after some calculations, it can be found that the $S'_k(\mathcal{N})$ secret key rate reduces to precisely

$$\underline{S}'_k(\mathcal{N}) = rS'_k(\mathcal{N})\frac{1}{(r+K_{in}-1)}. \tag{140}$$

Thus, the resulting $p_{err}\!\left(\underline{S}'_k(\mathcal{N})\right)$ error probability is [20]

$$p_{err}\!\left(\underline{S}'_k(\mathcal{N})\right) = \Pr\!\left(\log_2 \det\!\left(I_{K_{out}} + F(\mathbf{T}(\mathcal{N}))\frac{(\mathrm{SNR}')^*}{K_{in}} F(\mathbf{T}(\mathcal{N}))^\dagger\right) < \frac{S_k(r+K_{in}-1)}{r}P'(\mathcal{N})\right), \tag{141}$$

where $I_{K_{out}}$ is the $K_{out} \times K_{out}$ identity matrix.

Then, by exploiting a union bound averaged over the $\mathcal{S}$ statistics (see (61)) for each $\mathcal{N}_i$ [20–21], the $h_{K_{in}>K_{out}}$ optimal tradeoff curve is yielded as follows:

$$h_{K_{in}>K_{out}} : \delta_k(\varsigma_k) = 2(2 - \varsigma_k). \tag{142}$$

Assuming the situation $K_{in} \leq K_{out}$, some further results can also be derived.

By using (139) and the properties of the multidimensional manifold space $\mathcal{M}$ (see Theorem 1), and by averaging over the $\mathcal{S}$ statistics, the $h_{K_{in}\leq K_{out}}$ tradeoff function [20] without loss of generality is

$$h_{K_{in}\leq K_{out}} : \delta_k(\varsigma_k) = (i, (K_{in} - i) \cdot (K_{out} - i)), i = 0, \ldots, n_{\min}. \tag{143}$$

Putting the pieces together, for each function $h_{K_{in}>K_{out}}$ and $h_{K_{in}\leq K_{out}}$, the manifold extraction is optimized via the maximization of the $n_{\min}$ smallest singular values as $\lambda_{1\ldots n_{\min}}$, where for each $0 \leq \lambda_i \leq 2\sqrt{K_{in}}$ and are determined from $\mathbf{M}_j$ (see (132)).

Exploiting the argumentation of (110), the corresponding condition on $\left|\lambda_{1\ldots n_{\min}}\right|$ for the optimal tradeoff curve $h$ is precisely as

$$h : \max_{\forall i}\left|\lambda_{1\ldots n_{\min}}\right| > c^{\frac{1}{2}n_{\min}} \frac{1}{n_{\min}^{\frac{1}{2}n_{\min}} 2^{\frac{1}{2}S'_k(\mathcal{N}_i)}}, \tag{144}$$

for any constant $c > 0$.

The results for any $K_{in} > K_{out}$ and $K_{in} \leq K_{out}$ at $K_{in} = 2, K_{out} = 4$ are summarized in Fig. 5.

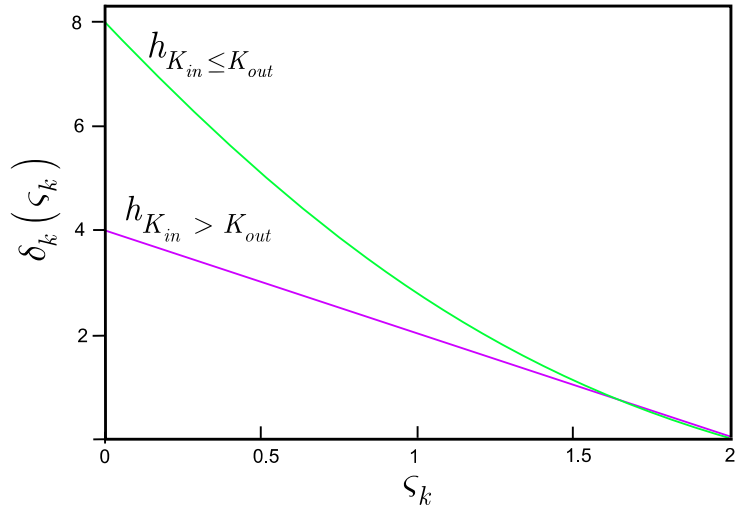

**Figure 5.** The optimal tradeoff curves $h_{K_{in}>K_{out}}$ for any $K_{in} > K_{out}$, and $h_{K_{in}\leq K_{out}}$ at $K_{in} = 2, K_{out} = 4$. The $h_{K_{in}>K_{out}}$ curve is maximized in $\delta_k(\varsigma_k) = 4$ at $\varsigma_k = 0$, and picks up the minimum $\delta_k(\varsigma_k) = 0$ at $\varsigma_k = 2$, for any $K_{in} > K_{out}$. For any $K_{in} \leq K_{out}$, the $h_{K_{in}\leq K_{out}}$ curve has the max. in $\delta_k(\varsigma_k) = K_{in}K_{out}$, $\varsigma_k = 0$, and the min. $\delta_k(\varsigma_k) = 0$ at $\varsigma_k = \min(K_{in}, K_{out})$.

■

# 5 Conclusions

The additional degree of freedom injected by the multicarrier transmission represents a significant resource to achieve performance improvements in CVQKD protocols. The proposed manifold extraction exploits those extra resources brought in by the multicarrier CVQKD modulation and is unavailable in a single-carrier CVQKD scheme. We introduced the term of multidimensional manifold extraction and proved that it can significantly improve the reliability of the phase space transmission. We demonstrated the results through the AMQD multicarrier modulation and extended it to the multiple-access multicarrier scenario through the AMQD-MQA scheme. We studied the potential of a multidimensional manifold space of multicarrier CVQKD and the optimized tradeoff curve between the manifold parameter and the additional degree of freedom ratio. The results confirm that the possibilities in a multicarrier CVQKD significantly exceed the single-carrier CVQKD scenario. The extra degrees of freedom allow the utilization of sophisticated optimization techniques for the aim of performance improvement. The available and efficiently exploitable extra resources have a crucial significance in experimental CVQKD, particularly in long-distance scenarios.

# Acknowledgements

The author would like to thank Professor Sandor Imre for useful discussions. This work was partially supported by the GOP-1.1.1-11-2012-0092 (*Secure quantum key distribution between two units on optical fiber network*) project sponsored by the EU and European Structural Fund, and by the COST Action MP1006.

# Supplemental Information

## S.1 Notations

The notations of the manuscript are summarized in Table S.1.

**Table S.1.** The summary of the notations.

| | |
|---|---|
| $Q(\cdot)$ | Gaussian tail function. |
| $rank(\cdot)$ | Rank function. |
| E | An event. |
| $i$ | Index for the $i$-th subcarrier Gaussian CV, $\lvert\phi_i\rangle = x_i + \mathrm{i}p_i$. |
| $j$ | Index for the $j$-th Gaussian single-carrier CV, $\lvert\varphi_j\rangle = x_j + \mathrm{i}p_j$. |
| $l$ | Number of Gaussian sub-channels $\mathcal{N}_i$ for the transmission of the Gaussian subcarriers. The overall number of the sub-channels is $n$. The remaining $n - l$ sub-channels do not transmit valuable information. |
| $(x_i, p_i)$ | Position and momentum quadratures of the $i$-th Gaussian subcarrier, $\lvert\phi_i\rangle = x_i + \mathrm{i}p_i$. |
| $(x'_i, p'_i)$ | Noisy position and momentum quadratures of Bob's $i$-th noisy subcarrier Gaussian CV, $\lvert\phi'_i\rangle = x'_i + \mathrm{i}p'_i$. |
| $(x_j, p_j)$ | Position and momentum quadratures of the $j$-th Gaussian single-carrier $\lvert\varphi_j\rangle = x_j + \mathrm{i}p_j$. |
| $(x'_j, p'_j)$ | Noisy position and momentum quadratures of Bob's $j$-th recovered single-carrier Gaussian CV $\lvert\varphi'_j\rangle = x'_j + \mathrm{i}p'_j$. |
| $x_{A,i}, p_{A,i}$ | Alice's quadratures in the transmission of the $i$-th subcarrier. |
| $\mathrm{SNR}_i$ | The SNR of the $i$-th Gaussian sub-channel $\mathcal{N}_i$, $\mathrm{SNR}_i = \frac{\sigma^2_{\omega_i}}{\sigma^2_{\mathcal{N}_i}}$. |
| $\text{SNR}'_i$ | The SNR of the $i$-th Gaussian sub-channel $\mathcal{N}_i$ in the SVD-assisted multicarrier transmission, $\text{SNR}'_i = \frac{\sigma^2_{\omega''_i}}{\sigma^2_{\mathcal{N}_i}}$. |
| $\text{SNR}$ | The SNR of the Gaussian channel $\mathcal{N}$, $\text{SNR} = \frac{\sigma^2_{\omega}}{\sigma^2_{\mathcal{N}}}$. |
| $\text{SNR}'$ | The SNR of the Gaussian channel $\mathcal{N}$ in an SVD-assisted protocol, $\text{SNR}' = \frac{\sigma^2_{\omega''}}{\sigma^2_{\mathcal{N}}}$. |
| $\text{SNR}^*_i$ | The SNR of the $i$-th Gaussian sub-channel $\mathcal{N}_i$ in a private transmission, $\text{SNR}^*_i = \frac{\sigma^2_{\omega_i}}{\sigma^2_{\mathcal{N}^*_i}}$. |
| $\left(\text{SNR}'_i\right)^*$ | The SNR of the $i$-th Gaussian sub-channel $\mathcal{N}_i$ in an SVD-assisted private transmission, $\left(\text{SNR}'_i\right)^* = \frac{\sigma^2_{\omega''_i}}{\sigma^2_{\mathcal{N}^*_i}}$. |
| $\text{SNR}^*$ | The SNR of the Gaussian channel $\mathcal{N}$ in a private transmission, $\text{SNR}^* = \frac{\sigma^2_{\omega}}{\sigma^2_{\mathcal{N}^*}}$. |
| $\left(\text{SNR}'\right)^*$ | The SNR of the Gaussian channel $\mathcal{N}$ in an SVD-assisted private transmission, $\left(\text{SNR}'\right)^* = \frac{\sigma^2_{\omega''}}{\sigma^2_{\mathcal{N}^*}}$. |
| $P(\mathcal{N}_i)$ | The private classical capacity of a Gaussian sub-channel $\mathcal{N}_i$. |
| $P'(\mathcal{N}_i)$ | The private classical capacity of a Gaussian sub-channel $\mathcal{N}_i$ at SVD-assistance. |
| $S(\mathcal{N})$, $S_k(\mathcal{N})$ | The secret key rate in a multicarrier setting, and the secret key rate of user $U_k$. |
| $S'(\mathcal{N})$, $S'_k(\mathcal{N}_i)$ | The secret key rate in an SVD-assisted multicarrier transmission, and the secret key rate of user $U_k$. In the manifold extraction these are fixed as $S'_k(\mathcal{N}) = \frac{\varsigma_k}{n_{\min}} P'(\mathcal{N})$, and $S'_k(\mathcal{N}_i) = \frac{\varsigma_{k,i}}{n_{\min}} P'(\mathcal{N}_i)$, respectively. |
| $\left|\delta_{1\ldots l}\right|$ | Product distance derived for the $l$ Gaussian sub-channels, $\left|\delta_{1\ldots l}\right|^2 > \left(c \frac{1}{l 2^{S'(\mathcal{N}_i)}}\right)^l$, for any constant $c > 0$ and secret key |

| | |
|---|---|
| | rate $S'(\mathcal{N}_i) > 0$ per $\mathcal{N}_i$. |
| $p_{err}$ | Error probability. |
| $\mathbf{p}_A$ | An $l$-dimensional random private codeword, $\mathbf{p}_A = (p_{A,1}, \ldots p_{A,l})^T$, where the $i$-th component $p_i$ is dedicated to $\mathcal{N}_i$. |
| $\mathcal{T}$ | Set of transmittance coefficients, such that for $\forall j$ of $\mathcal{T}: \left\|F(T_j(\mathcal{N}_j))\right\| = \min_{\forall i}\left\{\left\|F(T_i(\mathcal{N}_i))\right\|\right\}$, where $F(T_i(\mathcal{N}_i)) \geq \frac{1}{(\text{SNR}')^*}$. It refers to the worst-case scenario at which a $S'(\mathcal{N}) > 0$ nonzero secret key rate could exist. |
| $\mathcal{S}(\mathcal{N}_i)$ | A statistical averaging over the distribution of the $T_i(\mathcal{N}_i)$ transmittance coefficients. |
| $\chi^2_{2l}$ | Chi-square distribution with $2l$ degrees of freedom, has a density $f(x) = \frac{1}{(l-1)!}x^{l-1}e^{-x}$, where $x \geq 0$. |
| $\varsigma_k$ | Degree of freedom ratio of user $U_k$, $\varsigma_k = \frac{1}{P'(\mathcal{N})}S'_k(\mathcal{N})n_{\min}$. |
| $\delta_k$ | Manifold parameter, $\delta_k(\varsigma_k) = \lim_{(\text{SNR}')^* \to \infty} \frac{-\log_2 p_{err}(S'_k(\mathcal{N}))}{\frac{1}{n_{\min}}P'(\mathcal{N})}$. |
| $p_{err}^{single}$ | Error probability in a single-carrier transmission, $p_{err}^{single} = \frac{1}{\left((\text{SNR}')^*\right)^{\delta_{single}}}$, where $\delta_{single} = 1 - \varsigma$. |
| $p_{err}^{AMQD}$ | Error probability in a multicarrier transmission, $p_{err}^{AMQD} = \frac{1}{\left((\text{SNR}')^*\right)^{\delta_{AMQD}}}$, $\delta_{AMQD} = l(1 - \varsigma)$. |
| $\partial_k, \partial_{k,i}$ | Distance function for the phase space constellation $\mathcal{C}'_{\mathcal{S}}(\mathcal{N})$ and $\mathcal{C}'_{\mathcal{S}}(\mathcal{N}_i)$ of user $U_k$, $\partial_i = \frac{1}{2^{S'_k(\mathcal{N}_i)/2}}$, and $\partial_{k,i} = \frac{1}{2^{S'_k(\mathcal{N}_i)/2}}$. |
| $f$ | The optimal manifold-degree of freedom ratio tradeoff curve for a single-carrier transmission, $f: \delta_k(\varsigma_k) = Z(1 - \varsigma_k)$, where $0 < \varsigma_k \leq 1$. |
| $h$ | The optimal manifold-degree of freedom ratio tradeoff curve for multicarrier transmission, $f: \delta_k(\varsigma_k) = lZ(1 - \varsigma_k)$, |
| | where $0 < \varsigma_k \leq 1$, at $l$ sub-channels. |
| $r$ | Number of interfering sub-channels in an SVD-assisted multicarrier scenario. |
| $h(\mathcal{M})$ | The multidimensional optimal manifold-degree of freedom ratio tradeoff curve over the multidimensional manifold space $\mathcal{M}$. |
| $\mathcal{M}$ | Multidimensional manifold space, has dimension $\dim(\mathcal{M}) = K_{in}\varsigma_k + (K_{out} - \varsigma_k)\varsigma_k$. |
| $N_{\dim^\perp}$ | The number of dimensions orthogonal to $\mathcal{M}$ in the space of $S(F(\mathbf{T}(\mathcal{N})))$, $N_{\dim^\perp} = (K_{in} - \varsigma_k)(K_{out} - \varsigma_k)$. |
| $S(F(\mathbf{T}(\mathcal{N})))$ | The multidimensional space of $F(\mathbf{T}(\mathcal{N}))$, has dimension of $\dim(S(F(\mathbf{T}(\mathcal{N})))) = K_{in}K_{out}$. |
| $\lambda_i^2$ | The squared random singular values of $F(\mathbf{T}(\mathcal{N}))$. |
| $\mathbf{K}_o$ | An optimizing covariance matrix, defined as $\mathbf{K}_o = (\text{SNR}')^* \frac{I_{K_{in}}}{K_{in}}$, where $I_{K_{in}}$ is the $K_{in} \times K_{in}$ identity matrix. |
| $\mathbf{M}_j$ | The private codeword difference matrix, $\mathbf{M}_j = \frac{1}{\sqrt{(\text{SNR}')^*}}(\mathbf{p}_A - \mathbf{p}_B)$. |
| $\tilde{\lambda} = \min_{\forall i}(\lambda_i)$ | Smallest eigenvalue of the $\mathbf{M}_j$ private codeword difference matrix. |
| $\eta$ | A maximization criteria over the distribution of $F(\mathbf{T}(\mathcal{N}))$. |
| $S_1$, $S_2$ | Sets of singular operators $S_1 = \{F_1, U_2^{-1}\}$, $S_2 = \{U_1, U_2^{-1}\}$. |
| $F(\mathbf{T}) = U_2 \Gamma F_1^{-1}$, $F(\mathbf{T}) = U_2 \Gamma U_1^{-1}$ | The SVD of $F(\mathbf{T})$, where $F_1^{-1}, F_1 \in \mathbb{C}^{K_{in} \times K_{in}}$ and $U_2, U_2^{-1} \in \mathbb{C}^{K_{out} \times K_{out}}$ are unitary matrices, $K_{in}$ and $K_{out}$ refer to the number of sender and receiver users such that $K_{in} \leq K_{out}$, $F_1^{-1}F_1 = F_1 F_1^{-1} = I$, $U_2 U_2^{-1} = U_2^{-1} U_2 = I$, and $\Gamma \in \mathbb{R}$ is a diagonal matrix with non-negative real diagonal elements $\lambda_i$, $F(\mathbf{T}) = \sum_{n_{\min}} \lambda_i U_{2,i} F_{1,i}^{-1}$. |
| $\lambda_1 \geq \lambda_2 \geq \ldots \lambda_{n_{\min}}$ | The non-negative real diagonal elements of the diagonal matrix $\Gamma \in \mathbb{R}$, called the eigenchannels of |
| | $F(\mathbf{T}) = U_2 \Gamma F_1^{-1}$. |
| $\lambda_i^2$ | The $n_{\min}$ squared eigenchannels $\lambda_i^2$ are the eigenvalues of $F(\mathbf{T})F(\mathbf{T})^\dagger = U_2 \Gamma \Gamma^T U_2^{-1}$. |
| $n_{\min}$ | $n_{\min} = \min(K_{in}, K_{out})$, equals to the rank of $F(\mathbf{T})$, where $K_{in} \leq K_{out}$. |
| $\mathbf{s}$ | Stream matrix, $\mathbf{s} = (s_1, \ldots, s_{n_{\min}})^T \in \mathcal{CN}(0, \mathbf{K_s})$, defined by the unitary $F_1$ ($U_1$) applied on $\mathbf{z} \in \mathcal{CN}(0, \mathbf{K_z})$. |
| $s_i$ | A stream variable $s_i$ that identifies the CV state $|s_i\rangle$ in the phase space $\mathcal{S}$. Expressed as $|s_i'\rangle = \lambda_i U_{2,i} F_{1,i}^{-1} |s_i\rangle$, and $|\mathbf{s}'\rangle = F(\mathbf{T})\mathbf{s} = \sum_{n_{\min}} \lambda_i U_{2,i} F_{1,i}^{-1} |s_i\rangle$. |
| $U_2^{-1}(\gamma_i)$ | The Fourier-transformed eigenchannel interference, $U_2^{-1}(\gamma_i) \in \mathcal{CN}(0, \mathbf{K}_{U_2^{-1}(\gamma_i)})$, $\mathbf{K}_{U_2^{-1}(\gamma_i)} = \sigma_{\gamma_i}^2 = \mathbb{E}[|\gamma_i|^2]$, $|U_2^{-1}(\gamma_i)\rangle = U_2^{-1}\left(\sum_{j \neq i}^{n_{\min}} \lambda_j U_{2,j} F_{1,j}^{-1}\right)|s_j\rangle$. The variance $\sigma_\gamma^2 \to 0$, in the low-SNR regimes. |
| $z \in \mathcal{CN}(0, \sigma_z^2)$ | The variable of a single-carrier Gaussian CV state, $|\varphi_i\rangle \in \mathcal{S}$. Zero-mean, circular symmetric complex Gaussian random variable, $\sigma_z^2 = \mathbb{E}[|z|^2] = 2\sigma_{\omega_0}^2$, with i.i.d. zero mean, Gaussian random quadrature components $x, p \in \mathbb{N}(0, \sigma_{\omega_0}^2)$, where $\sigma_{\omega_0}^2$ is the variance. |
| $\Delta \in \mathcal{CN}(0, \sigma_\Delta^2)$ | The noise variable of the Gaussian channel $\mathcal{N}$, with i.i.d. zero-mean, Gaussian random noise components on the position and momentum quadratures $\Delta_x, \Delta_p \in \mathbb{N}(0, \sigma_\mathcal{N}^2)$, $\sigma_\Delta^2 = \mathbb{E}[|\Delta|^2] = 2\sigma_\mathcal{N}^2$. |
| $d \in \mathcal{CN}(0, \sigma_d^2)$ | The variable of a Gaussian subcarrier CV state, $|\phi_i\rangle \in \mathcal{S}$. Zero-mean, circular symmetric Gaussian random variable, $\sigma_d^2 = \mathbb{E}[|d|^2] = 2\sigma_\omega^2$, with i.i.d. zero mean, Gaussian random quadrature components $x_d, p_d \in \mathbb{N}(0, \sigma_\omega^2)$, where $\sigma_\omega^2$ |

| | |
|---|---|
| | is the modulation variance of the Gaussian subcarrier CV state. |
| $F^{-1}(\cdot) = \text{CVQFT}^{\dagger}(\cdot)$ | The inverse CVQFT transformation, applied by the encoder, continuous-variable unitary operation. |
| $F(\cdot) = \text{CVQFT}(\cdot)$ | The CVQFT transformation, applied by the decoder, continuous-variable unitary operation. |
| $F^{-1}(\cdot) = \text{IFFT}(\cdot)$ | Inverse FFT transform, applied by the encoder. |
| $\sigma^2_{\omega_0}$ | Single-carrier modulation variance. |
| $\sigma^2_{\omega} = \frac{1}{l}\sum_l \sigma^2_{\omega_i}$ | Multicarrier modulation variance. Average modulation variance of the $l$ Gaussian sub-channels $\mathcal{N}_i$. |
| $\begin{aligned}\lvert\phi_i\rangle &= \lvert\text{IFFT}(z_{k,i})\rangle \\ &= \lvert F^{-1}(z_{k,i})\rangle = \lvert d_i\rangle.\end{aligned}$ | The $i$-th Gaussian subcarrier CV of user $U_k$, where IFFT stands for the Inverse Fast Fourier Transform, $\lvert\phi_i\rangle \in \mathcal{S}$, $d_i \in \mathcal{CN}(0,\sigma^2_{d_i})$, $\sigma^2_{d_i} = \mathbb{E}\!\left[\lvert d_i\rvert^2\right]$, $d_i = x_{d_i} + \text{i}p_{d_i}$, $x_{d_i} \in \mathbb{N}(0,\sigma^2_{\omega_F})$, $p_{d_i} \in \mathbb{N}(0,\sigma^2_{\omega_F})$ are i.i.d. zero-mean Gaussian random quadrature components, and $\sigma^2_{\omega_F}$ is the variance of the Fourier transformed Gaussian state. |
| $\lvert\varphi_{k,i}\rangle = \text{CVQFT}(\lvert\phi_i\rangle)$ | The decoded single-carrier CV of user $U_k$ from the subcarrier CV, expressed as $F(\lvert d_i\rangle) = \lvert F(F^{-1}(z_{k,i}))\rangle = \lvert z_{k,i}\rangle$. |
| $\mathcal{N}$ | Gaussian quantum channel. |
| $\mathcal{N}_i, i=1,\ldots,n$ | Gaussian sub-channels. |
| $T(\mathcal{N})$ | Channel transmittance, normalized complex random variable, $T(\mathcal{N}) = \operatorname{Re}T(\mathcal{N}) + \text{i}\operatorname{Im}T(\mathcal{N}) \in \mathcal{C}$. The real part identifies the position quadrature transmission, the imaginary part identifies the transmittance of the position quadrature. |
| $T_i(\mathcal{N}_i)$ | Transmittance coefficient of Gaussian sub-channel $\mathcal{N}_i$, $T_i(\mathcal{N}_i) = \operatorname{Re}(T_i(\mathcal{N}_i)) + \text{i}\operatorname{Im}(T_i(\mathcal{N}_i)) \in \mathcal{C}$, quantifies the position and momentum quadrature transmission, with (normalized) real and imaginary parts $0 \leq \operatorname{Re}T_i(\mathcal{N}_i) \leq 1/\sqrt{2}$, $0 \leq \operatorname{Im}T_i(\mathcal{N}_i) \leq 1/\sqrt{2}$, where $\operatorname{Re}T_i(\mathcal{N}_i) = \operatorname{Im}T_i(\mathcal{N}_i)$. |
| $T_{Eve}$ | Eve's transmittance, $T_{Eve} = 1 - T(\mathcal{N})$. |
| $T_{Eve,i}$ | Eve's transmittance for the $i$-th subcarrier CV. |
| $\mathcal{A} \subseteq K$ | The subset of allocated users, $\mathcal{A} \subseteq K$. Only the allocated users can transmit information in a given (particularly the $j$-th) AMQD block. The cardinality of subset $\mathcal{A}$ is $|\mathcal{A}|$. |
| $U_k,\ k=1,\ldots,|\mathcal{A}|$ | An allocated user from subset $\mathcal{A} \subseteq K$. |
| $\mathbf{z} = \mathbf{x} + i\mathbf{p} = (z_1,\ldots,z_d)^T$ | A $d$-dimensional, zero-mean, circular symmetric complex random Gaussian vector that models $d$ Gaussian CV input states, $\mathcal{CN}(0,\mathbf{K_z})$, $\mathbf{K_z} = \mathbb{E}[\mathbf{zz}^\dagger]$, where $z_i = x_i + ip_i$, $\mathbf{x} = (x_1,\ldots,x_d)^T$, $\mathbf{p} = (p_1,\ldots,p_d)^T$, with $x_i \in \mathbb{N}(0,\sigma^2_{\omega_0})$, $p_i \in \mathbb{N}(0,\sigma^2_{\omega_0})$ i.i.d. zero-mean Gaussian random variables. |
| $\mathbf{d} = F^{-1}(\mathbf{z})$ | An $l$-dimensional, zero-mean, circular symmetric complex random Gaussian vector of the $l$ Gaussian subcarrier CVs, $\mathcal{CN}(0,\mathbf{K_d})$, $\mathbf{K_d} = \mathbb{E}[\mathbf{dd}^\dagger]$, $\mathbf{d} = (d_1,\ldots,d_l)^T$, $d_i = x_i + ip_i$, $x_i, p_i \in \mathbb{N}(0,\sigma^2_{\omega_F})$ are i.i.d. zero-mean Gaussian random variables, $\sigma^2_{\omega_F} = 1/\sigma^2_{\omega_0}$. The $i$-th component is $d_i \in \mathcal{CN}(0,\sigma^2_{d_i})$, $\sigma^2_{d_i} = \mathbb{E}[|d_i|^2]$. |
| $\mathbf{y}_k \in \mathcal{CN}(0,\mathbb{E}[\mathbf{y}_k \mathbf{y}_k^\dagger])$ | A $d$-dimensional zero-mean, circular symmetric complex Gaussian random vector. |
| $y_{k,m}$ | The $m$-th element of the $k$-th user's vector $\mathbf{y}_k$, expressed as $y_{k,m} = \sum_l F(T_i(\mathcal{N}_i))F(d_i) + F(\Delta_i)$. |
| $F(\mathbf{T}(\mathcal{N}))$ | Fourier transform of $\mathbf{T}(\mathcal{N}) = [T_1(\mathcal{N}_1)\ldots,T_l(\mathcal{N}_l)]^T \in \mathcal{C}^l$, the complex transmittance vector. |
| $F(\Delta)$ | Complex vector, expressed as $F(\Delta) = e^{\frac{-F(\Delta)^T \mathbf{K}_{F(\Delta)} F(\Delta)}{2}}$, with covariance matrix $\mathbf{K}_{F(\Delta)} = \mathbb{E}[F(\Delta)F(\Delta)^\dagger]$. |
| $\mathbf{y}[j]$ | AMQD block, $\mathbf{y}[j] = F(\mathbf{T}(\mathcal{N}))F(\mathbf{d})[j] + F(\Delta)[j]$. |
| $\tau = \|F(\mathbf{d})[j]\|^2$ | An exponentially distributed variable, with density $f(\tau) = (1/2\sigma^{2n}_\omega)e^{-\tau/2\sigma^2_\omega}$, $\mathbb{E}[\tau] \leq n2\sigma^2_\omega$. |
| $T_{Eve,i}$ | Eve's transmittance on the Gaussian sub-channel $\mathcal{N}_i$, $T_{Eve,i} = \operatorname{Re} T_{Eve,i} + \mathrm{i}\operatorname{Im} T_{Eve,i} \in \mathcal{C}$, $0 \leq \operatorname{Re} T_{Eve,i} \leq 1/\sqrt{2}$, $0 \leq \operatorname{Im} T_{Eve,i} \leq 1/\sqrt{2}$, $0 \leq \left|T_{Eve,i}\right|^2 < 1$. |
| $R_k$ | Transmission rate of user $U_k$. |
| $d_i$ | A $d_i$ subcarrier in an AMQD block. For subset $\mathcal{A} \subseteq K$ with $|\mathcal{A}|$ users and $l$ Gaussian sub-channels for the transmission, $d_i = \frac{1}{\sqrt{n}} \sum_{k=1}^{|\mathcal{A}|} z_k e^{\frac{-\mathrm{i}2\pi ik}{n}}, i = 1,\ldots,l$. |
| $\nu_{\min}$ | The $\min\{\nu_1,\ldots,\nu_l\}$ minimum of the $\nu_i$ sub-channel coefficients, where $\nu_i = \sigma_\mathcal{N}^2 \big/ \left|F\left(T_i\left(\mathcal{N}_i\right)\right)\right|^2$ and $\nu_i < \nu_{Eve}$. |
| $\sigma_\omega^2$ | Modulation variance, $\sigma_\omega^2 = \nu_{Eve} - \nu_{\min} \mathcal{G}(\delta)_{p(x)}$, where $\nu_{Eve} = \frac{1}{\lambda}$, $\lambda = \left|F\left(T_\mathcal{N}^*\right)\right|^2 = \frac{1}{n}\sum_{i=1}^n \left|\sum_{k=1}^n T_k^* e^{\frac{-\mathrm{i}2\pi ik}{n}}\right|^2$ and $T_\mathcal{N}^*$ is the expected transmittance of the Gaussian sub-channels under an optimal Gaussian collective attack. |
| $\nu_\kappa$ | Additional sub-channel coefficient for the correction of modulation imperfections. For an ideal Gaussian modulation, $\nu_\kappa = 0$, while for an arbitrary $p(x)$ distribution $\nu_\kappa = \nu_{\min}\left(1 - \mathcal{G}(\delta)_{p(x)}\right)$, where $\kappa = \frac{1}{\nu_{Eve} - \nu_{\min}\left(\mathcal{G}(\delta)_{p(x)} - 1\right)}$. |
| $\mathcal{N}_{U_k}[j] = [\mathcal{N}_1,\ldots,\mathcal{N}_s]^T$ | The set of $\mathcal{N}_i$ Gaussian sub-channels from the set of $l$ good sub-channels that transmit the $s$ subcarriers of user $U_k$ in the $j$-th AMQD block. |
| $\sigma_{\omega_i'}^2$ | The constant modulation variance $\sigma_{\omega_i'}^2$ for eigenchannel $\lambda_i$, evaluated as $\sigma_{\omega_i'}^2 = \mu - \left(\sigma_\mathcal{N}^2 \big/ \max_{n_{\min}} \lambda_i^2\right) = \frac{1}{n_{\min}}\sigma_{\omega'}^2$, with a total constraint $\sigma_{\omega'}^2 = \sum_{n_{\min}} \sigma_{\omega_i'}^2 = \frac{1}{l}\sum_l \sigma_{\omega_i}^2 = \sigma_\omega^2$. |
| $\sigma_{\omega''}^2$ | The modulation variance of the AMQD multicarrier transmission in the SVD environment. Expressed as $\sigma_{\omega''}^2 = \nu_{Eve} - \left(\sigma_\mathcal{N}^2 \big/ \max_{n_{\min}} \lambda_i^2\right)$, where $\lambda_i$ is the $i$-th eigen- |
| | channel of $F(\mathbf{T})$, $\max_{n_{\min}} \lambda_i^2$ is the largest eigenvalue of $F(\mathbf{T})F(\mathbf{T})^\dagger$, with a total constraint $\frac{1}{l}\sum_l \sigma_{\omega_i''}^2 = \sigma_{\omega''}^2 > \sigma_\omega^2$. |
| $S(F(\mathbf{T}))$ | The statistical model of $F(\mathbf{T})$ at a partial channel side information, $S(F(\mathbf{T})) = \xi_{K_{out}}^{-1} \Gamma \xi_{K_{in}}$, where $\xi_{K_{out}}^{-1}$ and $\xi_{K_{in}}$ are unitaries that formulate the input covariance matrix $\mathbf{K_s} = \xi_{K_{in}} \wp \xi_{K_{in}}^{-1}$, while $\wp$ is a diagonal matrix, $\mathbf{K_s} = Q diag\{\sigma_{\omega_1'}^2,...,\sigma_{\omega_{K_{in}}'}^2\}Q^\dagger$. |
| $\mathcal{C}_\mathcal{S}$ | Phase space constellation $\mathcal{C}_\mathcal{S}$. |
| $\mathcal{C}_\mathcal{S}^P(\mathcal{N})$ | Random phase space permutation constellation for the transmission of the Gaussian subcarriers, expressed as $\mathcal{C}_\mathcal{S}^P(\mathcal{N}) = \mathcal{C}_\mathcal{S}^P(\mathcal{N}_1),...,\mathcal{C}_\mathcal{S}^P(\mathcal{N}_l) = \left(\left|\phi_{1...d_{\mathcal{C}_\mathcal{S}^P(\mathcal{N}_1)}}\right\rangle, P_2\left|\phi_{1...d_{\mathcal{C}_\mathcal{S}^P(\mathcal{N}_2)}}\right\rangle,...,P_l\left|\phi_{1...d_{\mathcal{C}_\mathcal{S}^P(\mathcal{N}_l)}}\right\rangle\right)$, where $\left|\phi_i\right\rangle$ are the Gaussian subcarrier CVs, $P_i$, $i = 2,...,l$ is a random permutation operator, $d_{\mathcal{C}_\mathcal{S}(\mathcal{N}_i)} = d_{\mathcal{C}_\mathcal{S}(\mathcal{N}_j)}$ is the cardinality of $\mathcal{C}_\mathcal{S}(\mathcal{N}_i)$. The optimality function is $o(\mathcal{C}_\mathcal{S}(\mathcal{N})) = \sum_l \left(\nu_{Eve} - \left|\delta_i\right|^2\right)$. |
| $d_{\mathcal{C}_\mathcal{S}^P(\mathcal{N}_i)}$ | Cardinality of $\mathcal{C}_\mathcal{S}(\mathcal{N}_i)$. |
| $\delta_i$ | The normalized difference of two Gaussian subcarriers $d_{A,i}$ and $d_{B,i}$, $\delta_i = \frac{1}{\sqrt{\frac{\sigma_{\omega''}^2}{\sigma_\mathcal{N}^2}}}(d_{A,i} - d_{B,i})$. |
| $\partial$ | Difference function of Gaussian subcarriers (phase space symbols) $d_{A,i}$ and $d_{A,j}$ in constellations $\mathcal{C}_\mathcal{S}^P(\mathcal{N}_k)$, $k = 1,...,l$. For two Gaussian sub-channels $\mathcal{N}_1$ and $\mathcal{N}_2$, $\partial(\mathcal{C}_\mathcal{S}^P(\mathcal{N}_1)) = \min_{\forall d_{A,i}}(d_{A,i} - d_{A,j})$, $\partial(\mathcal{C}_\mathcal{S}^P(\mathcal{N}_2)) = \mho \cdot \partial(\mathcal{C}_\mathcal{S}^P(\mathcal{N}_1))$, where $\mho \geq 2$, $j \neq i$. |

## S.2 Abbreviations

| | |
|---|---|
| AMQD | Adaptive Multicarrier Quadrature Division |
| BS | Beam Splitter |
| CV | Continuous-Variable |
| CVQFT | Continuous-Variable Quantum Fourier Transform |
| CVQKD | Continuous-Variable Quantum Key Distribution |
| DV | Discrete Variable |
| FFT | Fast Fourier Transform |
| IFFT | Inverse Fast Fourier Transform |
| MQA | Multiuser Quadrature Allocation |
| QKD | Quantum Key Distribution |
| SNR | Signal to Noise Ratio |
| SNIR | Signal to Noise plus Interference Ratio |
| SVD | Singular Value Decomposition |